\documentclass[]{llncs}

\usepackage{xcolor,url}
\usepackage{graphicx,listings}
\usepackage{enumitem}
\usepackage{proof,xspace}
\usepackage{booktabs}
\usepackage{pgfplots}
\usepackage{wrapfig}

\newcommand{\secl}[1]{\mathit{sec} \ #1}
\newcommand{\siml}[3]{#1 \approx_{#2 l} #3}
\newcommand{\avx}{\forall \vec{X}{:}\vec{\tau}.~}
\newcommand{\mnot}{\mathit not}

\newcommand{\vsk}{\\[2ex]}

\newcommand{\lp}{logic programming\xspace}

\newcommand{\NF}{\emph{NF}\xspace}
\newcommand{\And}{\wedge}

\newcommand{\Imp}{\rightarrow}

\newcommand{\Bimp}{\leftrightarrow}

\newcommand{\defp}{\mbox{\textrm{def}}}

\renewcommand{\Pi}{\forall}  %%% all for pi

\newcommand{\wrt}{\textit{w.r.t}.~}
\newcommand{\via}{\textit{via}~}

\newcommand{\aprolog}{{\ensuremath \alpha}{Pro\-log}\xspace}
\newcommand{\acheck}{{\ensuremath \alpha}{Check}\xspace}
\newcommand{\hastype}{\mathrel{:}}

\newcommand{\gen}{\mathit{gen}}

\newcommand{\NE}{\emph{NE}\xspace}
\newcommand{\NEs}{\emph{NEs}\xspace}
\newcommand{\unitTy}{\mathbf{1}}
\newcommand{\pair}[1]{\langle #1 \rangle}
\newcommand{\comp}{\circ}

\newcommand{\constr}{{\mathcal K}}
\newcommand{\sat}[3]{#1;#2 \models #3}
\newcommand{\satgn}[1]{\Gamma;\constr \models #1}
\newcommand{\apf}[5]{#1;#2;#3 \stackrel{#4}{\too} #5}
\newcommand{\upf}[4]{#1;#2;#3 \Rightarrow #4}
\newcommand{\apfgdn}[2]{\apf{\Gamma}{\Delta}{\constr}{#1}{#2}}
\newcommand{\upfgdn}[1]{\Gamma;\Delta;\constr \Rightarrow #1}

\newbox\tempa
\newbox\tempb
\newdimen\tempc
\newbox\tempd
\newcommand{\mud}[1]{\hfil $\displaystyle{#1}$\hfil}
\newcommand{\rig}[1]{\hfil $\displaystyle{#1}$}

\newcommand{\lowerhalf}[1]{\hbox{\raise -0.8\baselineskip\hbox{#1}}}
\newcommand{\inruleanhelp}[3]{\setbox\tempa=\hbox{$\displaystyle{\mathstrut #2}$}%
                        \setbox\tempd=\hbox{$\; #3$}%
                        \setbox\tempb=\vbox{\halign{##\cr
        \mud{#1}\cr
        \noalign{\vskip\the\lineskip}%
        \noalign{\hrule height 0pt}%
        \rig{\vbox to 0pt{\vss\hbox to 0pt{\copy\tempd \hss}\vss}}\cr
        \noalign{\hrule}%
        \noalign{\vskip\the\lineskip}%
        \mud{\copy\tempa}\cr}}%
                      \tempc=\wd\tempb
                      \advance\tempc by \wd\tempa
                      \divide\tempc by 2 }

\newcommand{\neqt}{\mathit{neq}}
\newcommand{\nfr}{\mathit{nfr}}
\newcommand{\mnotd}{\ensuremath{\mathit{not^D}}}
\newcommand{\mnotg}{\ensuremath{\mathit{not^G}}}

\newcommand{\abs}[2]{{\ab{#1}{#2}}}
\newcommand{\conc}{\mathop{@}}
\newcommand{\unit}{{\ab{}}}
\newcommand{\ab}[1]{\langle #1 \rangle}

\newcommand{\andd}{\wedge}
\newcommand{\orr}{\vee}
\newcommand{\impp}{\supset}
\newcommand{\nott}{\neg}
\newcommand{\true}{\top}
\newcommand{\false}{\bot}
\newcommand{\eq}{\approx}
\newcommand{\eqt}[1]{\approx_{#1} \!}

\newcommand{\ent}{\mathrel{{:}{-}}}
\newcommand{\SB}[1]{[\![#1]\!]}
\newcommand{\idd}{\mathsf{id}}

\newcommand{\trueR}{{\top}R}

\newcommand{\impL}{{\impp}L}

\newcommand{\andL}{{\andd}L}
\newcommand{\andR}{{\andd}R}
\newcommand{\orL}{{\orr}L}
\newcommand{\orR}{{\orr}R}
\newcommand{\allL}{{\forall}L}

\newcommand{\exR}{{\exists}R}

\newcommand{\hyp}[2][]{\infer[#1]{#2}{}}

\newcommand{\newR}{{\new}R}

\newcommand{\new}[1][]{\reflectbox{\sf{#1{}N}}}

\newcommand{\fresh}{\mathrel{\#}}
\newcommand{\fresht}[1]{\mathrel{\#}_{#1}}
\newcommand{\Aa}{\name{a}}
\newcommand{\Ab}{\name{b}}
\newcommand{\Ac}{\name{c}}

\newcommand{\name}[1]{\mathsf{#1}}
\newcommand{\too}{\longrightarrow}
\newcommand{\act}{\cdot}
\newcommand{\tran}[2]{(#1~#2)}

\newcommand{\swap}[3]{(#1~#2)\act#3}

\newcommand{\labelFig}[1]{\label{fig:#1}}
\newcommand{\refFig}[1]{Figure~\ref{fig:#1}}

\newcommand{\btab}{\begin{tabbing}}
\newcommand{\etab}{\end{tabbing}}

\newcommand{\bit}{\begin{itemize}}
\newcommand{\enit}{\end{itemize}}

\newbox\tempa
\newbox\tempb
\newdimen\tempc
\newbox\tempd

\def\mud#1{\hfil $\displaystyle{#1}$\hfil}
\def\rig#1{\hfil $\displaystyle{#1}$}

\def\inruleanhelp#1#2#3{\setbox\tempa=\hbox{$\displaystyle{\mathstrut #2}$}%
                        \setbox\tempd=\hbox{$\; #3$}%
                        \setbox\tempb=\vbox{\halign{##\cr
        \mud{#1}\cr
        \noalign{\vskip\the\lineskip}%
        \noalign{\hrule height 0pt}%
        \rig{\vbox to 0pt{\vss\hbox to 0pt{\copy\tempd \hss}\vss}}\cr
        \noalign{\hrule}%
        \noalign{\vskip\the\lineskip}%
        \mud{\copy\tempa}\cr}}%
                      \tempc=\wd\tempb
                      \advance\tempc by \wd\tempa
                      \divide\tempc by 2 }

\def\inrulean#1#2#3{{\inruleanhelp{#1}{#2}{#3}%
                     \hbox to \wd\tempa{\hss \box\tempb \hss}}}

\def\lowerhalf#1{\hbox{\raise -0.8\baselineskip\hbox{#1}}}

\long\def\ednote#1{\footnote{[{\it #1\/}]}\message{ednote!}}
\newenvironment{metanote}{\begin{quote}\message{note!}[\begingroup\it}%
                         {\endgroup]\end{quote}}
\long\def\ignore#1{}

\newcommand{\sstlc}{\emph{Stlc}\xspace}

\def\bnfas{\mathrel{::=}}

\def\oftp{\mathord{:}}
\def\hastype{\mathrel{:}}

\begin{document}
\title{Advances in Property-Based Testing for \aprolog}

\author{James Cheney\inst{1}, Alberto Momigliano\inst{2}, Matteo Pessina\inst{2}}
\institute{University of Edinburgh
\email{jcheney@inf.ed.ac.uk}
  \and
Universit\`{a} degli Studi
    di Milano \\\email{momigliano@di.unimi.it, matteo.pessina3@studenti.unimi.it}}
% \begin{keywords}
% nominal logic programming, property-based testing, counterexample search,
%   negation elimination
%   \end{keywords}
%\def\titlerunning{Run}
%\def\authorrunning{Cheney,  Momigliano \& Pessina}

\maketitle

\begin{abstract}
  $\alpha$Check is a light-weight property-based testing tool built on
  top of $\alpha$Prolog, a logic programming language based on nominal
  logic.  $\alpha$Prolog is particularly suited to the validation of
  the meta-theory of formal systems, for example correctness of
  compiler translations involving name-binding, alpha-equivalence and
  capture-avoiding substitution. In this paper we describe an
  alternative to the negation elimination algorithm underlying \acheck
   that substantially improves its effectiveness.
  To substantiate this claim we compare the checker performances
  w.r.t.\ two of its main competitors in the logical framework niche,
  namely the QuickCheck/Nitpick combination offered by Isabelle/HOL
  and the random testing facility in PLT-Redex.
\end{abstract}
% -------------------
%\input{intro}
\section{Introduction}

Formal compiler verification has come a long way from McCarthy and
Painter's ``Correctness of a Compiler for Arithmetic Expression''
(1967), as witnessed by the success of \emph{CompCert} and subsequent
projects~\cite{Leroy09,CompCertTSO}. However outstanding these
achievements are, they are not a magic wand for every-day compiler
writers: not only CompCert was designed with verification in mind,
whereby the implementation and the verification were a single process,
but there are only a few dozen people in the world able and willing
to carry out such an endeavour. By verification, CompCert means the
preservation of certain simulation relations between source,
intermediate and target code; however, the translations involved are
relatively simple compared to those employed by modern optimizing compilers.
% ensuring that nothing bad occurs, but
% not necessarily that something good does occur, that is, performance
% have indeed been optimized. 
Despite some initial work~\cite{Aspinall07,Breitner}, handling 
more realistic optimizations seems even harder, e.g.\ the verification
of the \emph{call
  arity} analysis and transformation in the Glasgow Haskell Compiler (GHC):
\begin{small}
  \begin{quote}
    % \begin{footnotesize}
    ``The [Nominal] Isabelle development corresponding to this paper,
    including the definition of the syntax and the semantics, contains
    roughly 12,000 lines of code with 1,200 lemmas (many small, some
    large) in 75 theories, created over the course of 9 months'' (page
    11,~\cite{Breitner}).
    % \end{footnotesize}
  \end{quote}
\end{small}

For the rest of us, hence, it is back to compiler testing, which is
basically synonymous with passing a hand-written fixed validation
suite. This is not completely satisfactory, as the coverage of those
tests is difficult to assess and because, being fixed, these suites
will not uncover new bugs. In the last few years, \emph{randomized
  differential testing}~\cite{McKeeman98} has been suggested in
combination with automatic generation of (expressive) test programs,
most notably for C compilers with the \emph{Csmith} tool~\cite{Csmith}
and to a lesser extent for GHC~\cite{Palka2011}. The oracle is
\emph{comparison checking}: %~\cite{JaramilloGS02}:
Csmith feeds randomly generated programs to several compilers and
flags the minority one(s), that is, those reporting different
outputs from the majority of the other compilers under
test, as incorrect. Similarly, the outcome of GHC on a random program with or
without an optimization enabled is compared.

%\smallskip
\emph{Property-based testing}, as pioneered by
QuickCheck~\cite{claessen00icfp}, seems to leverage the automatic
generation of test cases with the use of \emph{logical specifications} (the
properties), making validation possible not only in a differential
way, but internally, \wrt (an abstraction of) the behavior of the
source and intermediate code. In fact, compiler
verification/validation is a prominent example of the more general
field of  verification of the \emph{meta-theory} of formal systems.
For many classes of (typically) shallow bugs, a tool that
automatically finds counterexamples can be surprisingly effective and
can complement formal proof attempts by warning when the property we
wish to prove has easily-found counterexamples. The beauty of such
\emph{meta-theory model checking} is that, compared to other general
forms of system validation, the properties that should hold are
already given by means of the theorems that the calculus under study
is supposed to satisfy.  Of course, those need to be fine tuned for
testing to be effective, but we are mostly free of the thorny issue of
specification/invariant generation.

In fact, such tools are now gaining  traction in the field of
semantics engineering, see in particular the QuickCheck/Nitpick
combination offered in Isabelle/HOL~\cite{BlanchetteBN11} and random
testing in PLT-Redex~\cite{Klein12}. However, a particular
dimension to validating for example optimizations in a compiler such
as GHC, whose intermediate language is a variant of the
polymorphically typed $\lambda$-calculus, is a correct, simple and
effective handling of \emph{binding signatures} and associated notions
such as $\alpha$-equivalence and capture avoiding substitutions. A
small but not insignificant part of the success of the CompCert
project is due to not having to deal with any notion of
binder\footnote{X.~Leroy, personal communication. In fact,
  the encoding % of the $\lambda$-calculus 
in~\cite{LeroyG09} does not
  respect $\alpha$-equivalence, nor does it implement substitutions in
  a capture avoiding way.}. The ability to encode possibly
non-algorithmic relations (such as typing) in a declarative way would
also be a plus.

The nominal logic programming language \aprolog~\cite{cheney08toplas}
offers all those
facilities. % and has been used in significant case studies
Additionally, it was among the first to propose a form of property
based testing for language specifications with the \emph{\acheck}
tool~\cite{CheneyM07}. In contrast to QuickCheck/Nitpick and PLT
Redex, our approach supports binding syntax directly and uses logic
programming to perform \emph{exhaustive symbolic} search for
counterexamples. Systems lacking this kind of support may end up with
ineffective testing capabilities or requiring an additional amount of
coding, which needs to be duplicated in every case study:
\begin{quote}
  \begin{small}
    ``Redex offers little support for handling binding constructs in
    object languages. It provides a generic function for obtaining a
    fresh variable, but no help in defining capture-avoiding
    substitution or $\alpha$-equivalence [ \dots] In one case [ \dots]
    managing binders constitutes a significant portion of the overall
    time spent [ \dots] Generators derived from grammars [ \dots] require substantial
    massaging to achieve high test coverage. This deficiency is
    particularly pressing in the case of typed object languages, where
    the massaging code almost duplicates the specification of the type
    system'' (page 5,~\cite{Klein12}).
  \end{small}
\end{quote}

\acheck extends \aprolog with tools for searching for counterexamples,
that is,  substitutions that makes the
antecedent of a specification true and the conclusion false. In logic
programming terms this means fixing a notion of \emph{negation}. To
begin with, \acheck adopted the infamous \emph{negation-as-failure} (NF)
operation, ``which put pains thousandfold upon the'' logic
programmers. As many good things in life, its conceptual simplicity
and efficiency is marred by significant problems:
\begin{itemize}[noitemsep]
\item the lack of an agreed intended semantics against which to carry
  a soundness proof: this concern is significant because the semantics
  of negation as failure has not yet been investigated for nominal
  logic programming;
\item even assuming such a semantics, we know that \NF is unsound for
  non-ground goals; hence all free variables must be instantiated
  before solving the negated conclusion. This is obviously
  exponentially expensive in an exhaustive search setting and may prevent
  optimizations by goal reordering.
\end{itemize}

%\smallskip
To remedy this \acheck also offered \emph{negation elimination}
(NE)~\cite{Bar90,Momigliano00}, a source-to-source transformation that
replaces negated subgoals to calls to equivalent positively defined
predicates. \NE by-passes the previous issues arising for \NF since,
in the absence of local (existential) variables, it yields an ordinary
($\alpha$)Prolog program, whose intended model is included in the
complement of the model of the source program.  In particular, it avoids
the expensive term generation step needed for \NF, it has been proved
correct, and it may open up other opportunities for
optimization. Unfortunately, in the experiments reported in our
initial implementation of \acheck~\cite{CheneyM07}, \NE turned out to
be slower than \NF.

Perhaps to the reader's chagrin, this paper does not tackle the
validation of compiler optimizations (yet). Rather, it lays the
foundations by:
\begin{enumerate}[noitemsep]
\item describing an alternative implementation of negation
  elimination, dubbed \NEs ---``s'' for simplified: this improves
  significantly over the performance of \NE as described
  in~\cite{CheneyM07} by producing negative programs that are
  equivalent, but much more succinct, 
 so much as to make the method competitive \wrt  \NF;
\item and by evaluating our checker
in comparison with some of its competitors in the logical framework
niche, namely QuickCheck/Nitpick~\cite{BlanchetteBN11} and
PLT-Redex~\cite{Klein12}.  To the best of our knowledge, this is
the first time any of these three tools have been compared experimentally.
\end{enumerate}
%of this paper is twofold: first,  Second,  % and the
  % prppf-theoretic cointermodel-search offered by
  % Bedwyr~\cite{BaeldeGMNT07}

\smallskip In the next section we give a tutorial presentation of the
tool and move then to the formal description of the logical engine
(Section~\ref{sec:core}). In Section~\ref{sec:checker}, we detail the
\NEs algorithm and its implementation, whereas Section~\ref{sec:comp}
carries out the promised comparison on two case studies, a
prototypical $\lambda$-calculus with lists and a basic type system for
secure information flow.  The Appendix contains  some formal
notions~(\ref{sec:app}) used in Section~\ref{sec:core} and additional experiments~(\ref{sec:quick}). 

The sources for \aprolog and \acheck can be
found at \url{https://github.com/aprolog-lang/aprolog}. Supplementary
material, including the full listing of the case studies presented
here  are available at~\cite{TAPapp}. We assume some familiarity with
logic programming.

% ------------------
 \section{A Brief Tour of  \acheck}
\label{sec:tut}

We specify the formal systems and the  properties we wish to check
 as Horn logic programs in
\aprolog~\cite{cheney08toplas}, a logic programming
language based on \emph{nominal logic}, a first-order theory
axiomatizing names and name-binding introduced by Pitts~\cite{pitts03ic}.
% and based on Gabbay and Pitts' swapping-based approach to binding
% syntax~\cite{gabbay02fac}.

In \aprolog, there are several built-in types, functions, and relations
with special behavior.  There are distinguished \emph{name types} that
 are populated with infinitely many \emph{name constants}.  In
program text, a lower-case identifier is considered to be a name
constant by default if it has not already been declared as something else.  Names can be used in
\emph{abstractions}, written \verb|a\M| in programs, considered equal
up to $\alpha$-renaming of the bound name.  Thus, where one writes
$\lambda x. M$, $\forall x. M$, etc.\ in a paper exposition, in \aprolog
one writes \verb|lam(x\M)|, \verb|forall(x\M)|, etc.  In addition, the
\emph{freshness} relation \verb$a # t$ holds between a name \verb$a$
and a term \verb$t$ that does not contain a free occurrence of
\verb$a$.  Thus, $x \not\in FV(t)$ is written in \aprolog
as \verb|x # t|; in particular, if $t$ is also a name then freshness
is name-inequality.  For convenience, \aprolog provides a
function-definition syntax, but this is just translated to an
equivalent (but more verbose) relational implementation via \emph{flattening}.

Horn logic programs over these operations suffice to define a wide
variety of object languages, type systems, and operational semantics in
a convenient way. To give a feel of the interaction with the checker,
here we encode a simply-typed $\lambda$-calculus augmented with
constructors for integers and lists, following the PLT-Redex benchmark
\texttt{sltk.lists.rkt} from
\url{http://docs.racket-lang.org/redex/benchmark.html}, which we will
examine more deeply in Section~\ref{ssec:plt}.  The language is
formally declared as follows:

\[
\begin{array}{llcl}
\mbox{Types}& A,B & \bnfas & int \mid ilist \mid A \rightarrow B  \\
\mbox{Terms} & M & \bnfas &  x \mid \lambda {x\oftp A}.\ {M} \mid
{M_1}\ {M_2} \mid c\mid err\\
\mbox{Constants} & c & \bnfas &  n \mid nil\mid cons\mid hd \mid tl\\
\mbox{Values} & V & \bnfas &  c \mid \lambda {x\oftp A}.\ {M} \mid
cons\ V \mid
cons\ V_1 \ V_2
\end{array}
\]

We start (see the top of Figure~\ref{fig:example}) by declaring the
syntax of terms, constants and types, while we carve out values
\via an appropriate predicate. A similar predicate \texttt{is\_err}
characterizes the threading in the operational semantics of the
\emph{err} expression, used to model run time errors such as taking
the head of an empty list.
%\cite{HarperPFPL}. % and of a typing context as a list of
% variable-types pairs.
% We assume some familiarity with nominal logic
% and just mention that \verb!lam(x\var(x),intTy)! is concrete syntax
% for $\lambda x\oftp int.\ x$.
% the lambda terms denoting the identity function.
%\begin{figure}
%  \centering
%   \begin{small}
% \begin{verbatim}
% ty: type.         id: name_type.  exp: type.             cst: type.
% intTy: ty.        listTy: ty.     funTy: (ty,ty) -> ty. 
% var: id -> exp.   c: cst -> exp.  app: (exp,exp) -> exp. err: exp.
% lam: (id\exp,ty) -> exp.          cons: cst.             hd: cst.        
% tl: cst.          nil: cst.       toInt: int -> cst.  
% \end{verbatim}
%   \end{small}
%   \caption{Syntax of STLC with booleans}
%   \label{fig:stlc}
% \end{figure}

  \begin{figure}[tb!]
\label{fig:tut}
    \centering
 %   \begin{small}
\begin{verbatim}
ty: type.         
intTy: ty.          funTy: (ty,ty) -> ty.     listTy: ty.     
cst: type.
toInt: int -> cst.  nil: cst.  cons: cst.  hd: cst.  tl: cst.
id: name_type.  
exp: type.             
var: id -> exp.     lam: (id\exp,ty) -> exp.  app: (exp,exp) -> exp.  
c: cst -> exp.      err: exp.

type ctx = [(id,ty)]. 

pred tc (ctx,exp,ty).
tc(_,err,T).
tc(_,c(C),T)                              :- tcf(C) = T.
tc([(X,T)|G],var(X),T).
tc([(Y,_)|G],var(X),T)                    :- X # Y, tc(G,var(X),T).
tc(G,app(M,N),U)                          :- tc(G,M,funTy(T,U)), tc(G,N,T).
tc(G,lam(x\M,T),funTy(T,U))               :- x # G, tc([(x,T) |G],M,U).

pred step(exp,exp).
step(app(c(hd),app(app(c(cons),V),VS)),V) :- value(V), value(VS).
step(app(c(tl),app(app(c(cons),V),VS)),VS):- value(V), value(VS).
step(app(lam(x\M,T),V), subst(M,x,V))     :- value(V).
step(app(M1,M2),app(M1',M2))              :- step(M1,M1').
step(app(V1,M2),app(M1,M2'))              :- value(V1), step(M2,M2').

pred is_err(exp).
is_err(err).
is_err(app(c(hd),c(nil)))).
is_err(app(c(tl),c(nil))).
is_err(app(E1,E2))                        :- is_err(E1).
is_err(app(V1,E2))                        :- value(V1), is_err(E2).
\end{verbatim}
%    \end{small}
    \caption{Encoding of the example calculus in \aprolog}\label{fig:example}
  \end{figure}

  We follow this up (see the remainder of Figure~\ref{fig:example})
  with the static semantics (predicate \texttt{tc}) and
  dynamic semantics (one-step reduction predicate \texttt{step}), where we omit the judgments for
  the \texttt{value} predicate and \texttt{subst} function, which are analogous to the
  ones in~\cite{CheneyM07}. Note that \emph{err} has any type and
  constants are typed \via a table \texttt{tcf}, also omitted.
% \begin{figure}
%   \centering
%   \begin{small}
% \begin{verbatim}
% type ctx = [(id,ty)]. 

% pred tc (ctx,exp,ty).
% tc(_,err,T).
% tc(_,c(C),T)                                 :- tcf(C) = T.
% tc([(X,T)|G],var(X),T).
% tc([(Y,_)|G],var(X),T)                       :- X # Y, tc(G,var(X),T).
% tc(G,app(M,N),U)                             :- tc(G,M,funTy(T,U)), tc(G,N,T).
% tc(G,lam(x\M,T),funTy(T,U))                  :- x # G, tc([(x,T) |G],M,U).

% pred step(exp,exp).
% step(app(c(hd),app(app(c(cons),V),VS)),V)    :- value(V), value(VS).
% step(app(c(tl),app(app(c(cons),V),VS)),VS)   :- value(V), value(VS).
% step(app(lam(x\M,T),V), subst(M,x,V))        :- value(V).
% step(app(M1,M2),app(M1',M2))                 :- step(M1,M1').
% step(app(V1,M2),app(M1,M2'))                 :- value(V1), step(M2,M2').

% pred is_err(exp).
% is_err(err).
% is_err(app(c(hd),c(nil)))).
% is_err(app(c(tl),c(nil))).
% is_err(app(E1,E2))                           :- is_err(E1).
% is_err(app(V1,E2))                           :- value(V1), is_err(E2).
% \end{verbatim}
% \end{small}
% \caption{Static and dynamic semantics of STLCb}
% \end{figure}

Horn clauses can also be used as specifications of desired program
properties of such an encoding, including basic lemmas concerning
substitution as well as main theorems such as preservation, progress,
and type soundness.  This is realized \via checking
\emph{directives}
\begin{verbatim}
#check "spec" n : H1, ..., Hn => A.
\end{verbatim}
where \verb|spec| is a label naming the property, \verb|n| is a
parameter that bounds the search space, and \verb$H1$ through
\verb$Hn$ and \verb$A$ are atomic formulas describing the
preconditions and conclusion of the property.  As with program
clauses, the specification formula is implicitly universally
quantified. 
  Following the PLT-Redex development, we
  concentrate here only on checking that that preservation
  and progress hold.%  We remark,  as we have
  % shown in \cite{CheneyM07}, that it is often worthwhile
  % to check also auxiliary lemmas, e.g.\ that substitution is
  % functional or that weakening holds for the typing judgments, since
  % this can help uncover bugs that may not show up if only the main
  % statement is tested.
 
\begin{small}
\begin{verbatim}
#check "pres" 7 : tc([],E,T), step(E,E')  => tc([],E',T).
#check "prog" 7 : tc([],E,T) => progress(E).
\end{verbatim}
\end{small}
Here, \texttt{progress} is a
predicate encoding the property of ``being either a value, an error,
or able to make a step''.  The tool will not find any counterexample,
because, well, those properties are (hopefully) true of the given
setup. Now, let us insert a typo that
swaps the range and domain types of the function in the application
rule, which now reads:
\begin{small}
\begin{verbatim}
tc(G,app(M,N),U) :- tc(G,M,funTy(T,U)), tc(G,N,U). % was funTy(U,T)
\end{verbatim}
\end{small}
Does any property become false?  The checker returns immediately with
this counterexample to progress:
\begin{small}
\begin{verbatim}
E = app(c(hd),c(toInt(N))) 
T = intTy 
\end{verbatim}
\end{small}
This is abstract syntax for ${hd\ n}$, an expression erroneously
well-typed and obviously stuck.
Preservation meets a similar fate: $(\lambda x{:} T \Imp int.\ x \
err) \ n$ steps to an ill-typed term.
\begin{small}
\begin{verbatim}
E = app(lam(x\app(var(x),err),funTy(T,intTy)),c(toInt(N))) 
E' = app(c(toInt(N)),err) 
T = intTy 
\end{verbatim}
\end{small}

%%% Local Variables:
%%% mode: latex
%%% TeX-master: t
%%% End:

% ------------------
% \input{core}
%\newcommand{\aprolog}{{\ensuremath \alpha}{Pro\-log}\xspace}

\newcommand{\cred}[1]{\color{red}{\ensuremath{#1}}}

\section{The Core Language}
\label{sec:core}
In this section we give the essential notions concerning the core
syntax, to which the surface syntax used in the previous section desugars,
and semantics of \aprolog programs.

An \aprolog \emph{signature} % $\Sigma =
% (\Sigma_D,\Sigma_N,\Sigma_P,\Sigma_F)$
is composed by sets $\Sigma_D$
and $\Sigma_N$ of, respectively, base  types $\delta$, which
includes a type $o$ of \emph{propositions}, and name types $\nu$; a
collection $\Sigma_P$ of \emph{predicate symbols} $p : \tau \to o $
and one $\Sigma_F$ of \emph{function symbol} declarations $f : \tau
\to \delta$. Types $\tau$ are formed as specified by the following
grammar:
\begin{eqnarray*}
  \tau &::=&  \delta \mid \tau \times \tau' \mid\unitTy \mid \nu \mid
  \abs{\nu}\tau 
\end{eqnarray*}
where $\delta \in \Sigma_D$ and $\nu \in \Sigma_N$ and $\unitTy$ is
the unit type.  Given a signature, the language of \emph{terms} is
defined over sets $V = \{X,Y,Z,\ldots\}$ of logical variables and sets
$A = \{\Aa,\Ab,\ldots\}$ of names:
\begin{eqnarray*}
  t,u &::=& \Aa \mid \pi \act X \mid \unit \mid \pair{t,u} \mid
  \abs{\Aa}{t}\mid f(t) \\
  \pi &::=& \idd \mid\tran{\Aa}{\Ab}\comp \pi
\end{eqnarray*}
where $\pi$ are permutations, which we omit in case $\idd \act X$,
$\unit$ is unit, $\pair{t,u}$ is a pair and $\abs{\Aa}{t}$ is the
abstract syntax for name-abstraction.  The result
of applying the permutation $\pi$ (considered as a function) to
$\Aa$ is written $\pi(\Aa)$.  Typing for these terms is standard, with
the main novelty being that name-abstractions $\abs{\Aa}{t}$ have
abstraction types $\abs{\nu}\tau$ provided $\Aa : \nu$ and $t :\tau$.

% The effect of a permutation $\pi$ on a name is defined as follows:
% \[\begin{array}{rcl}
% \idd (\Aa) &=& \Aa\\
% (\tran{\Aa}{\Ab} \comp \pi)(\Ac) &=& \left\{\begin{array}{ll}
% \Ab & \pi(\Ac) = \Aa\\
% \Aa & \pi(\Ac) = \Ab\\
% \pi(\Ac) & \pi(\Ac) \notin \{\Aa,\Ab\}
% \end{array}\right.
% \end{array}\]

% The swapping operation is defined on \emph{ground} terms in
% the following way:
% \[\begin{array}{rclcrclc}
%   \pi \act \unit &= &\unit&\qquad&
%   \pi \act f(t) &=& f(\pi \act t)\\
%   \pi \act \pair{t,u} &=& \pair{\pi \act t, \pi \act u}& \qquad &
%   \pi \act \Aa &= &\pi(\Aa)\\
%   \pi \act \abs{\Aa}{t} &= &\abs{\pi \act \Aa}{\pi \act t}
% \end{array}\]

The \emph{freshness} ($s \fresht{\tau} u$) and \emph{equality} ($t\eqt{\tau} u$)
constraints, where $s$ is a term of some name type $\nu$, are the
new features provided by nominal logic.  The former
relation is defined on ground terms by the following
inference rules, where $f : \tau \to
\delta\in\Sigma_F$:
\[\begin{array}{c}
  \infer{\Aa \fresht\nu \Ab}{\Aa \neq \Ab}\quad
  \infer{\Aa \fresht{\unitTy} \unit}{}\quad
  \infer{\Aa \fresht\delta f(t)}{\Aa \fresht\tau t}\quad
  \infer{\Aa \fresht{\tau_1 \times \tau_2} \pair{t_1,t_2}}
{\Aa \fresht{\tau_1} t_1 & \Aa \fresht{\tau_2} t_2}
\quad   \infer{\Aa \fresht{\abs{\nu'}{\tau}} \abs{\Ab}{t}}
{\Aa \fresht{\nu'} \Ab & \Aa \fresht{\tau} t}\quad
  \infer{\Aa \fresht{\abs{\nu'}{\tau}} \abs{\Aa}{t}}{}
\end{array}\]
In the same way we define the equality relation, which identifies
terms modulo $\alpha$-equivalence, where $\swap{\Aa}{\Ab}{u}$ denotes
\emph{swapping} two names in a term:
\[\begin{array}{c}
  \hyp{\Aa \eqt{\nu} \Aa}\qquad
  \hyp{\unit \eqt{\unitTy}  \unit}\qquad
  \infer{\pair{t_1,t_2} \eqt{\tau_1 \times \tau_2} \pair{u_1,u_2}}
{t_1 \eqt{\tau_1} u_1 & t_2 \eqt{\tau_2} u_2}\qquad
  \infer{f(t) \eqt{\delta} f(u)}{t \eqt{\tau} u}
\vsk
  \infer{\abs{\Aa}{t} \eqt{\abs{\nu}{\tau}} \abs{\Ab}{u}}
{ \Aa \eqt\nu \Ab & t \eqt\tau u }\qquad
  \infer{\abs{\Aa}{t} \eqt{\abs{\nu}{\tau}} \abs{\Ab}{u}}
{\Aa \fresht\nu \Ab\quad \Aa \fresht\nu u & t \eqt\tau \swap{\Aa}{\Ab}{u}}
\end{array}\]
% We omit the type subscript when it is
% clear from the context.

Given a signature,
\emph{goals} $G$ and \emph{program clauses} $D$ have the following form:
\begin{eqnarray*}
  A &::=& t \eq u \mid t \fresh u\\
  G &::=& \false \mid \true \mid A \mid p(t) \mid G \andd G' \mid G \orr G' 
  \mid \exists X{:}\tau.~G \mid \new \Aa{:}\nu.~G \ \cred {\mid \forall^*X{:}\tau.~G}\\ 
  D &::=& \true \mid p(t) \mid D \andd D' \mid G \impp D \mid \forall
          X:\tau.~D \  \cred{\mid \false \mid D \orr D'}  
\end{eqnarray*}
The productions shown in black yield a fragment of nominal logic
called $\new$-goal clauses~\cite{cheney08toplas}, for which resolution
based on nominal unification is sound and complete.  This is in
contrast to the general case where the more complicated
\emph{equivariant unification} problem must be
solved~\cite{cheney10jar}. We rely on the fact that $D$ formulas in a
program $\Delta$ can always be normalized to sets of clauses of the
form ${\avx}G \impp p(t)$, denoted $\defp(p,\Delta)$. The
\emph{fresh-name} quantifier $\new$, firstly introduced
in~\cite{pitts03ic}, quantifies over names not occurring in a formula
(or in the values of its variables).%  It can be defined in terms of
% freshness $\fresh$, that is the formula $\new \Aa{:}\nu.~\phi$ is
% logically equivalent to
% $\exists A{:}\nu.~A \fresh \vec{X} \andd \phi $, where $\vec{X}$ are
% the free variables appearing in $\phi$.  Here we take
% $\new$-quantified formulas as primitive.
%% omitted for space
The extensions shown in red here in the language BNF (and in its
proof-theoretic semantics in~\refFig{uapf-aprolog}) instead are
constructs brought in from the negation elimination procedure
(Section~\ref{subsec:ne}) and which will not appear in any source
programs. In particular, an unusual feature is the \emph{extensional}
universal quantifier  $\forall^*$~\cite{Harland93}.  Differently from the
\emph{intensional} universal quantifier $\forall$, for which
$\forall X{:}\tau.~G$ holds if and only if $G[x/X]$ holds, where $x$
is an eigenvariable representing any terms of type $\tau$,
$\forall^*X{:}\tau.~G$  succeeds if and only if $G[t/X]$ does for
\emph{every} ground term of type $\tau$.

\emph{Constraints} are $G$-formulas of the following form:
\[
C ::= \top \mid t \eq u \mid t \fresh u \mid C \andd C' \mid \exists X{:}\tau.~C
\mid \new \Aa{:}\nu.~C
\]
We write $\constr$ for a set of constraints and $\Gamma$ for a context
keeping track of the types of variables and names.  Constraint-solving is
modeled by the judgment \mbox{$\satgn{C}$}, which holds if for all
maps $\theta$ from variables in $ \Gamma$ to ground terms
if $\theta \models \constr$ then  $\theta \models C$. The
latter notion of satisfiability is standard, modulo handling of names:
for example
$\theta \models \new \Aa{:}\nu.~C$ iff \mbox{for some} $\Ab $ fresh
for $
\theta$ and $C$,  $ \theta \models C[\Ab/\Aa]$.
%  Let \theta \models C[\Ab/\Aa]$.
%  Let
% $\theta$ be a valuation that is a function from variables to ground
% terms. We say that $\theta$ matches $\Gamma$ ($\theta :
% \Gamma$) if $\theta(X) : \Gamma(X)$ for each $X$, and all 
% implicit freshness constraints in $\Gamma$ are satisfied, namely if
% $\Gamma = \Gamma_1,X:\tau,\Gamma_2\#\Aa{:}\nu,\Gamma_3$ then $a \fresh
% \theta(X)$. is defined as follows:

\begin{figure}[t]
  \[
  \begin{array}{c}
     \infer[con]
    {\upfgdn{A}}
    {\satgn{A}}
    \qquad
    \infer[\andR]{\upfgdn{G_1 \andd G_2}}
    {\upfgdn{G_1} & \upfgdn{G_2}}
                    \medskip\\
    \infer[\orR_i]{\upfgdn{G_1 \orr G_2}}
    {\upfgdn{G_i}}
    \qquad
    \infer[\exR]{\upfgdn{\exists X{:}\tau.~G}}
    { \sat{\Gamma}{\constr}{\exists X{:}\tau.~C} &
      \upf{\Gamma,X{:}\tau}{\Delta}{\constr,C}{G} }  
                                                   \medskip\\
    \infer[\newR]{\upfgdn{\new\Aa{:}\nu.~G}}
    {\sat{\Gamma}{\constr}{\new \Aa{:}\nu.~C} &
      \upf{\Gamma\#\Aa{:}\nu}{\Delta}{\constr,C}{G} } 
                                                \medskip\\
    \infer[\trueR]{\upfgdn{\top}}{} 
    \qquad
    \infer[sel]{\upfgdn{Q}}{\apfgdn{D}{Q} & D \in \Delta}\medskip\\ 
\infer[\cred{\forall^*\omega}]{\cred{\upfgdn{\forall^{*} X{:}\tau.~G}}}{
  \cred{\bigwedge \{\upf{\Gamma,X{:}\tau}{\Delta}{\constr,C}{G} \mid
  \satgn{\exists X{:}\tau.~C}\}}} \\
    \dotfill\\[1em]
    \infer[hyp]{\apfgdn{p(t)}{p(u)}}{\satgn{t\eq u}}
    \qquad
    \infer[\andL_i]{\apfgdn{D_1 \andd D_2}{Q}}{\apfgdn{D_i}{Q}}
    \medskip\\
    \infer[\impL]{\apfgdn{G \impp D}{Q}}
    {\apfgdn{D}{Q} & \upfgdn{G}}
    \medskip\\
    \infer[\allL]
    {\apfgdn{\forall X{:}\tau.~D}{Q}}
    {\sat{\Gamma}{\constr}{\exists X{:}\tau.~C} &
        \apf{\Gamma,X{:}\tau}{\Delta}{\constr,C}{D}{Q} } \medskip\\
\infer[\cred{\false}L]
  {\cred{\apfgdn{\false}{Q}}}
   {}
  \qquad
  \infer[\cred{\orL}]
  {\cred{\apfgdn{D_1 \orr D_2}{Q}}}
  {\cred{\apfgdn{D_1}{Q}} &\cred{\apfgdn{D_2}{Q}}}
  \end{array}
  \]
  \caption{Proof search semantics of \aprolog
    programs}\labelFig{uapf-aprolog}
  % \[
  %   \begin{array}{c}
  %     \infer[back]{\upfgdn{p(u)}}
  %     {\sat{\Gamma}{\constr}{\exists \vec{X}{:}\vec{\tau}.~\vec C \wedge t\eq
  %     u } & \upf{\Gamma,\vec{X}{:}\vec{\tau}}{\Delta}{\constr \wedge
  %           \vec C}{G} & (\forall \vec{X}{:}\vec{\tau}.~G \impp p(t))\in\Delta}
  %   \end{array}
  % \]
  % \caption{Backchaining rule}\label{fig:back}
\end{figure}

We can describe an idealized interpreter for \aprolog with the
``amalgamated'' proof-theoretic semantics introduced
in~\cite{cheney08toplas} and inspired by similar techniques stemming
from CLP~\cite{leach01tplp} --- see \refFig{uapf-aprolog}, sporting
two kind of judgments, goal-directed proof search $\upfgdn{G}$ and
focused proof search $\apfgdn{D}{Q}$.
% \begin{itemize}
% \item 
% \item program clause-directed (or 
% \end{itemize}
% where $\Delta$ is a set of clauses,  
This semantics allows us to concentrate on the
high-level proof search issues, without requiring  to introduce or
manage low-level operational details concerning constraint solving.
We refer the reader to~\cite{cheney08toplas} for more explanation and
ways to make those judgments operational.
% semantics of \aprolog programs, which is a partitioned set of inference
% rules based on two kind of judgment: 
% \begin{itemize}
% \item goal-directed (or uniform proof search) $\upfgdn{G}$
% \item program clause-directed (or focused proof search)$\apfgdn{D}{Q}$
% \end{itemize}
% where $\Delta$ is a set of clauses, $\constr$ is a set of constraints
% and colon and semicolon are just syntactic separators.
% This formulation follows the semantics introduced in \cite{cheney08toplas},
% stemming from Constraint Logic Programming (CLP)\cite{leach01tplp}. 
Note that the rule $\forall^*\omega$ says that goals of the form
$\forall^* X{:}\tau.G$ can be proved if
$\upf{\Gamma,X{:}\tau}{\Delta}{\constr,C}{G}$ is provable for every
constraint $C$ such that $\satgn{\exists X{:}\tau.~C}$ holds.  Since
this is hardly practical, the number of candidate constraints $C$
being infinite, we approximate it by modifying the interpreter so as
to perform a form of case analysis: at every stage, as dictated by the
type of the quantified variable, we can either instantiate $X$ by
performing a one-layer type-driven case distinction and further recur
to expose the next layer by introducing new $\forall^*$ quantifiers,
or we can break the recursion by instantiation with an eigenvariable.
%parameter. %viewing $\forall^*$ as {generic} quantification.
% For convenience we provide a backchaining rule for normalized
% clauses (\ref{fig:back}), but it is derivable from the ones
% already defined.

\section{Specification Checking}
\label{sec:checker}

Informally, \verb|#check| specifications correspond to
specification formulas of the form
\begin{equation}\label{eq:typical-spec}
 \new \vec{\Aa}. \forall \vec{X}.~G \impp A
\end{equation}
where $G$ is a goal and $A$ an atomic formula (including equality and
freshness constraints).  Since the $\new$-quantifier is self-dual, the
negation of  (\ref{eq:typical-spec}) is of the form
$\new \vec{\Aa}. \exists{\vec{X}}. G \andd \neg A$.  A \emph{(finite)
  counterexample} is a closed substitution $\theta$ providing values
for $\vec{X}$ such that $\theta(G)$ is derivable, but the conclusion
$\theta(A)$ is not. 
 Since we live in a \lp world, the choice of what
we mean by ``not holding'' is crucial, as we must choose an
appropriate notion of \emph{negation}. 

In \acheck the reference implementation reads negation as \emph{finite
failure} (\textit{not}):
\begin{eqnarray}
\label{eqn:nf}
 \new \vec{\Aa}. \exists \vec{X}{:}\vec{\tau}.~G \andd 
 \gen\SB{\vec\tau}(\vec X) \andd not(A)
\end{eqnarray}
where $\gen\SB{\vec\tau}$ are type-indexed predicates that
\emph{exhaustively} enumerate the (ground) inhabitants of
$\vec\tau$. For example, $ \gen\SB{\mathtt{ty}}$ yields the predicate:
\begin{small}
\begin{verbatim}
gen_ty(intTy).          gen_ty(listTy). 
gen_ty(funTy(T1,T2)) :- gen_ty(T1), gen_ty(T2).
\end{verbatim}
\end{small}
 % The operator \texttt{not} is Prolog's (in)famous \NF.
A check such as~(\ref{eqn:nf}) can simply be executed as a goal in
the \aprolog interpreter, using the number of resolution steps
permitted to solve each subgoal as a bound on the search space.  This
method, combined with a complete search strategy such as iterative
deepening, will find a counterexample, if one exists. 
%As we have mentioned in the Introduction, 
This realization of
specification checking is simple and  effective, while not
escaping the traditional problems associated with such an operational
notion of negation.

%  We define the \emph{bounded model checking}
% problem for such programs and properties as follows: given a resource
% bound (\eg a bound on the number of inference steps needed), decide
% whether a counterexample can be derived using the given resources, and
% if so, compute such a counterexample.

% \subsection{Negation-as-Failure}
% \label{ssec:nf}

% Checking via Negation-as-Failure is realized in \aprolog
% transforming a specification in the following formula:
% \begin{equation}
%  \new \vec{\Aa}. \exists \vec{X}{:}\vec{\tau}.~H_1 \andd \ldots \andd
%  H_n \andd
%  \gen\SB{\tau}(X_1) \andd \cdots \andd \gen\SB{\tau}(X_n)\andd not(A)
% \end{equation}

% Note that this approach is \emph{derivation-first}: it generates all
% ``finished'' derivations of the hypothesis $\vec{H}$ up to a given
% bound, considers all the sufficiently ground instantiations of
% variables in each up to the depth bound, and then tests if the
% conclusion finitely fails for the resulting substitutions.

% A peculiar characteristic implemented in \NF is an \emph{ad-hoc}
% resource bound on the test conclusion, often referred in the thesis as
% ``the augmented bound''.  In the current implementation it is hard
% coded to $3 * bound + 10$.  This is motivated by the fact the often
% the derivation of the ground conclusion requires a deeper search
% space.

\subsection{Negation Elimination}
\label{subsec:ne}

Negation Elimination~\cite{Bar90,Momigliano00} is a source-to-source
transformation that replaces negated subgoals with calls to a combination
of equivalent positively defined predicates.  In the absence of local
(existential) variables, \NE yields an ordinary ($\alpha$)Prolog
program, whose intended model is included in the
complement of the model of the source program.  In
other terms, a predicate and its complement are mutually
\emph{exclusive}. \emph{Exhaustivity}, that is whether a program and
its complement coincide with the Herbrand base of  the program's
signature may or may not hold, depending on
the decidability of the predicate in question; nevertheless, this property,
though desirable, is neither frequent nor necessary in a model
checking context.  When local variables are present, the derived
positivized program  features the \emph{extensional} universal quantifier
presented in the previous section.

% In the context of specification checking, \NE has many advantages \wrt
% \NF. The most importants is the ability to work with open terms, thus
% skipping the expensive procedures that make open terms ground needed
% in \NF.
%  Moreover it
% avoids complex semantics and implementation issues arising for \NF.

The generation of complementary predicates can be split into two
phases: \emph{term complementation} and \emph{clause complementation}.

\paragraph{Term complementation}
A cause of atomic goal failure is when its arguments do not unify with
any of the program clause heads in its definition.  The idea is then
to generate the complement of the term structure in each clause head
by constructing a set of terms that differ in at least one
position. However, and similarly to the higher-order logic case, the
complement of a nominal term containing free or bound names cannot
be represented by a \emph{finite} set of nominal terms. For our
application nonetheless, we can pre-process clauses so that the standard
complementation algorithm for (linear) first order terms applies~\cite{Lassez87}.
This forces  terms in source clause heads to be linear and free of
names (including swapping and abstractions), by
replacing them with logical variables, and, in case they occurred in
abstractions, by constraining them in the clause body by a \emph{concretion}
to a fresh variable.  A concretion, written $t@a$, is the elimination
form for abstractions and %  does not require to be taken as a primitive
% since it 
can be implemented by translating a goal $G$ with an
occurrence of $[t@a]$ (notation $G[t@a]$) to
$\exists X.t \eq \ab{a}X \andd G[X]$.  For example, the clause for
typing lambdas is normalized as: \\
% \begin{small}
% \begin{verbatim}
\texttt{tc(G,lam(M,T),funTy(T,U)):- new x. tc([(x,T) |G],M@x,U).}
% \end{verbatim}
% \end{small}

Hence, we can use a type-directed version of first-order term
complementation, $ \mnot\SB{\tau} : \tau \to \tau\ \mathit{set}$ and
prove its correctness in term of \emph{exclusivity}
following~\cite{Bar90,Momigliano03}: the intersection of the set of
ground instances of a term and its complement is
empty. \emph{Exhaustivity} also holds, but will not be needed. The
definition of $ \mnot\SB{\tau}$ is in the appendix~\ref{sec:app}, but we offer the
following example:
\begin{small}
  \begin{eqnarray*}
    \lefteqn{\mnot\SB{\mathtt{exp}}(\mathtt{app(c(hd),\_ ) })=}\\
 &&       \{\mathtt{lam(\_,\_),                                             err,c(\_),var(\_),app(c(tl),\_),app(c(nil),\_),
    app(c(toInt(\_)),\_)}, \\
&& \mathtt{app(var(\_),\_), app(err,\_),app(lam(\_,\_),\_), app(app(\_,\_),\_)} \}
  \end{eqnarray*}
\end{small}

% (fig.~\ref{fig:complterms})
% , where $f
% : \tau \rightarrow \delta$.

% \begin{figure}
%  \begin{small}
%   \begin{eqnarray*}
%    \mnot\SB{\tau} &:&  \tau \to o\ \mathit{set}\\
%     \mnot\SB\tau(t) & = & \emptyset\qquad\qquad\qquad\mbox{when } \tau\in\{
%     \unitTy,\nu, \abs{\nu}\tau \} \mbox{ or $t$ is a variable}\\
%     \mnot\SB{\tau_1 \times \tau_2}(t_1,t_2) & = & 
% \{ (s_1,\_) \mid s_1\in\mnot\SB{\tau_1}(t_1)\}\cup\{ (\_,s_2) \mid
% s_s\in\mnot\SB{\tau_2}(t_2)\}\\
% \mnot\SB\delta(f(t)) & = & \{
% g(\_)\mid g\in\Sigma, g \hastype \sigma\Imp \delta, f\ \neq g\} \cup \{f(s)\mid s\in\mnot\SB\tau(t) \}
%   \end{eqnarray*}
% \end{small}
% \caption{Rules for term complementing}
% \label{fig:complterms}
% \end{figure}

% The correctness of the algorithm for term complementation is  
% analogous to previous
% results~\cite{Bar90,Momigliano03},%  can be stated in the following
% constraint-conscious way, as required by the proof of the main
% Theorem~\ref{thm:exclu}:
% \begin{lemma}[Term Exclusivity] 
% \label{le:excluT}

% Let $\constr$ be consistent, $ s \in \mnot\SB\tau(t)$,
% $FV(u)\subseteq\Gamma$ and $FV(s,t)\subseteq\vec X$.  It is not the
% case that both $\sat{\Gamma}{\constr}{\exists \vec X {:} \vec \tau.~u
%   \eq t %\land \vec C
% }$ and $\sat{\Gamma}{\constr}{\exists \vec X {:} \vec \tau.~ u \eq
%   s % \land \vec D
% }$.
% \end{lemma}

\paragraph{Clause complementation}

\begin{figure}[t]
%\begin{eqnarray*}
\[
\begin{array}{rclrrl}
\mnotg(\true) &=& \false & \mnotd(\true) &=& \false\\
\mnotg(\false) &=& \true &  \mnotd(\false) &=& \true \\
\mnotg(p(t)) &=& p^\nott(t)& \mnotd(G \impp p(t)) &=& \bigwedge
                                                      \{\forall (p^\nott(u)) \mid
                                u \in\mnot\SB\tau(t)\}\andd \mbox{} \\ 
                &&& & & 
                    (\mnotg(G) \impp p^\nott(t))\\
\mnotg(t \eq_\tau u) &=& \neqt\SB{\tau}(t,u)\\
\mnotg(a \fresh_{\tau} u) &=& \nfr\SB{\nu,\tau}(a,u)\\
\mnotg(G \andd G') &=& \mnotg(G) \orr \mnotg(G') &   \mnotd(D \andd D') &=& \mnotd(D) \orr \mnotd(D')\\
\mnotg(G \orr G') &=& \mnotg(G) \andd \mnotg(G')&  \mnotd(D \orr D') &=& \mnotd(D) \andd \mnotd(D')\\
\mnotg(\forall^*X{:}\tau.~G) &=& \exists X{:}\tau.~\mnotg(G) &
                                                               \mnotd(\forall X{:}\tau.~D) &=& \forall X{:}\tau.~\mnotd(D)\\
\mnotg(\exists X{:}\tau.~G) &=& \forall^*X{:}\tau.~\mnotg(G)\\
\mnotg(\new \Aa{:}\nu.~G) &=& \new \Aa{:}\nu.~\mnotg(G) & \mnotd({\Delta}) &=& \mnotd(\defp(p,\Delta))
\end{array}
\]
\caption{Negation of a goal and of clause}\labelFig{notgoal}
%\labelFig{notclause}
\end{figure}

The idea of the clause complementation algorithm is to compute the
complement of each head of a predicate definition using term
complementation, while clause bodies are negated pushing negation
inwards until atoms are reached and replaced by their complement and
the negation of constraints is computed.  The contributions (in fact a
disjunction) of each of the original clauses are finally
merged. % via unification.
The whole procedure can be seen as a negation normal form procedure,
which is consistent with the operational semantics of the language.

The first ingredient is complementing the equality and freshness
constraints, yielding  ($\alpha$-)inequality  $\neqt\SB{\tau}$ and
non-freshness  $\nfr\SB{\nu,\delta}$: 
we implement these using type-directed code generation within the
\aprolog interpreter and refer again to the appendix~\cite{TAPapp} for their
 generic definition.

\refFig{notgoal} shows goal and clause complementation: most cases of
the former, \via 
the $\mnotg$ function, are intuitive, being classical tautologies.
Note that the self-duality of the $\new$-quantifier allows goal
negation to be applied recursively.  Complementing existential goals
is where we introduce \emph{extensional} quantification and invoke its
proof-theory.

Clause complementation is where things get interesting and differ from
the previous algorithm~\cite{CheneyM07}. The complement of a clause
$G \impp p(t)$ must contain a ``factual'' part, built \via term
complementation, motivating failure due to clash with (some term in)
the head. We obtain the rest by negating the body with $\mnotg
(G)$.
We take clause complementation \emph{definition-wise}, that is the
negation of a program is the conjunction of the negation of all its
predicate definitions.  An example may help: negating the typing
clauses for constants and application (\texttt{tc} from
Fig.~\ref{fig:tut}) produces the following disjunction:
\begin{small}
\begin{verbatim}
(not_tc(_,err,_) /\ not_tc(_,var(_),_) /\ not_tc(_,app(_,_),_) /\
 not_tc(_,lam(_,_),_) /\ not_tc(_,c(C),T):- neq(tcf(C), T))
 \/
(not_tc(_,err,_) /\ not_tc(_,var(_),_) /\ not_tc(_,c(_),_) /\ 
 not_tc(_,lam(_,_),_) /\
 not_tc(G,app(M,N),U):- forall* T. not_tc(G,M,funTy(T,U)) /\
 not_tc(G,app(M,N),U):- forall* T. not_tc(G,N,T))
\end{verbatim}
\end{small}
Notwithstanding the top-level disjunction, we are \emph{not}
committing to any form of disjunctive \lp: the key observation is that
`$\orr$' can be restricted to a program constructor \emph{inside} a
predicate definition; therefore it can be eliminated by simulating
unification in the definition:
$$ ( G_{1} \impp Q_{1}) \orr (G_{2} \impp Q_{2})\equiv \theta( G_1\And
G_2\impp Q_{1})$$
where $\theta = \mbox{mgu}(Q_{1},Q_{2})$.  Because $\orr$ is
commutative and associative we can perform this merging operation in
any order. However, as with many bottom-up operations, merging tends
to produce a lot of redundancies in terms of clauses that are
instances of each other. We have implemented \emph{backward} and
\emph{forward} subsumption~\cite{Loveland94}, by using an extension of
the \aprolog interpreter itself to check entailment between newly
generated clauses and the current database (and vice-versa). Despite
the fact that this subsumption check is \emph{partial}, because the
current unification algorithm does not handle equivariant unification
with mixed prefixes~\cite{Miller92} and extensional
quantification~\cite{cheney10jar}, it makes all the difference: the
\verb|not_is_err| predicate definition decreases from an unacceptable
$128$ clauses to a much more reasonable $18$.  The final definition of
\texttt{not\_tc} follows, where we (as in Prolog) use the semicolon as concrete syntax for
disjunction in the body:
\begin{small}
\begin{verbatim}
not_tc(_,c(C),T)             :- neq_ty(tcf(C),T).
not_tc([],var(_),_).
not_tc([(X,T)|G],var(X'),T') :- (neq_ty(T,T'); fresh_id(X,X')), 
                                not_tc(G,var(X'),T').
not_tc(G,app(M,N),U)         :- forall* T:ty. not_tc(G,M,funTy(T,U)); 
                                              not_tc(G,N,T).
not_tc(G,app(M,N),listTy)    :- forall* T:ty. not_tc(G,M,funTy(T,listTy)); 
                                              not_tc(G,N,T).
not_tc(G,app(M,N),intTy)     :- forall* T:ty. not_tc(G,M,funTy(T,intTy)); 
                                              not_tc(G,N,T).
not_tc(_,lam(_),listTy).
not_tc(_,lam(_),intTy).
not_tc(G,lam(M,T),funTy(T,U)):- new x:id. not_tc([(x,T)|G],M@x,U).
\end{verbatim}
\end{small}
Regardless of the presence of two subsumed clauses in the \texttt{app}
case that our approach failed to detect, it is a big improvement in
comparison to the $38$ clauses generated by the previous
algorithm~\cite{CheneyM07}. And in exhaustive search, every clause
counts.

Having synthesized the negation of the \texttt{tc} predicate, \acheck
will use it internally while
searching, for instance in the preservation check, for
$$\exists E.\exists T.~\mathtt{tc([],\mathit{E},\mathit{T}), step(\mathit{E},\mathit{E}'),
  not\_tc([],\mathit{E}',\mathit{T})}$$

Soundness of clause complementation is  crucial for the
purpose of model checking; we again express it in terms of
exclusivity. The proof follows the lines of~\cite{Momigliano00}.
\begin{theorem}[Exclusivity] 
\label{thm:exclu}
 Let $\constr$ be consistent. It is not the case that:
 \begin{itemize}[noitemsep]
 \item  $\upfgdn{G}$ and
   $\upf{\Gamma}{\mnotd(\Delta)}{\constr} {\mnotg(G)}$;
 \item  $\apfgdn{D}{Q}$ and
   $\apf{\Gamma}{\mnotd(\Delta)}{\constr} {\mnotd(D)}{\mnotg(Q)}$.
 \end{itemize}
\end{theorem}
%
%%% Local Variables:
%%% mode: latex
%%% TeX-master: "tap"
%%% End:

% -------------------
%\input{experiment}
\section{Case Studies}
\label{sec:comp}

We have chosen as case studies here the \sstlc benchmark suite,
introduced in Section~\ref{sec:tut}, and an encoding of the Volpano et
al.\ security type system~\cite{Volpano1996}, as suggested in
\cite{BlanchetteITP10}.  For the sake of space, we report \emph{at the
  same time} our comparison between the various forms of negation, in
particular \NEs vs.~\NE, \emph{and} the other systems of reference,
accordingly, PLT-Redex and Nitpick.

\emph{PLT-Redex}~\cite{PLTbook} is an executable DSL for mechanizing
semantic models built on top of \emph{DrRacket}.  Redex has been the
first environment to adopt the idea of random testing a la QuickCheck
for validating the meta-theory of object languages, with significant
success~\cite{Klein12}. As we have mentioned, the main drawbacks
are the lack of support for binders and low coverage
of % : variables are just another non-terminal and
% they are handled in an ad hoc way. A generic substitution
% (meta)function is provided but it has to be tweaked to respect binding
% occurrences.
%The tool provides naive 
test generators stemming from grammar definitions. The user is
therefore required to write her own generators, a task which tends to be
demanding.

The system where proofs and disproofs are best integrated is arguably
Isabelle/HOL~\cite{BlanchetteBN11}. % , which offers a combination of
% random, exhaustive and symbolic testing~\cite{Bulwahn12}.  
 In the
appendix~\ref{sec:app} we report some comparison with its version of
QuickCheck, but here we concentrate on
\emph{Nitpick}~\cite{BlanchetteITP10}, a higher-order model finder in
the \emph{Alloy} lineage supporting (co)inductive definitions. Nitpick
works translating a significant fragment of HOL into first-order
relational logic and then invoking Alloy's SAT-based model
enumerator. The tool has been used effectively in several case
studies, most notably weak memory models for
C++~\cite{BlanchetteWBOS11}.  It would be natural to couple
Isabelle/HOL's QuickCheck and/or Nitpick's capabilities with
\emph{Nominal} Isabelle~\cite{urban12lmcs}, but this would require
strengthening the latter's support for computation with names, permutations and
abstract syntax modulo $\alpha$-conversion. So, at the time of
writing, \acheck is unique as a model checker for binding signatures
and specifications.

All test have been performed under Ubuntu 15.4 on a Intel Core i7 CPU
870, 2.93GHz with 8GB RAM\@. We time-out the computation when it
exceeds 200 seconds. We report $0$ when the time is $<$0.01.  These
tests must be taken with a lot of salt: not only is our tool under
active development but the comparison with the other systems is only
roughly indicative, having to factor differences between logic and
functional programming (PLT-Redex), as well as the sheer scale and scope
of counter-examples search in a system such as Isabelle/HOL\@.

\subsection{Head-to-Head with PLT-Redex}
\label{ssec:plt}

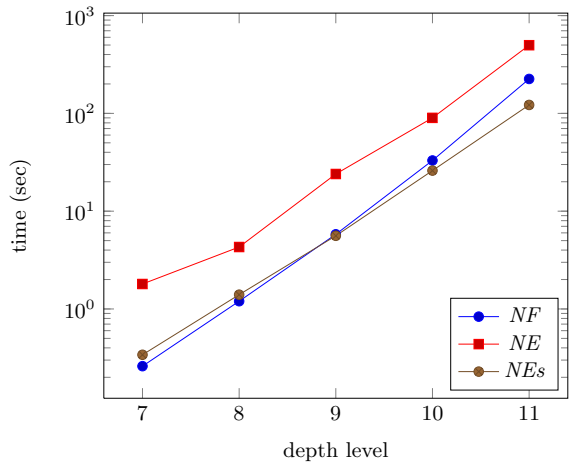
\begin{figure}[t]
%\vspace{-20pt}
 \centering
\begin{tikzpicture} [scale=0.90]
\begin{axis}[
    ymode=log,
    xlabel={depth level},
    ylabel={time (sec)},
    legend entries={\NF,\NE,\NEs},
    legend style={at={(0.98,0.02)},anchor=south east}
]
% NF
\addplot table{
7 0.26
8 1.2
9 5.8
10 33
11 225
};
% NE
\addplot table{
7 1.8
8 4.3
9 24
10 90
11 498
};
% NEs
\addplot table{
7 0.34
8 1.4
9 5.6
10 26
11 122
};
\end{axis}
\end{tikzpicture}
%\vspace{-10pt}
\caption{Loglinear-plot of TESS on prog theorem}
%\vspace{-20pt}
\label{fig:log_expl_stlc}
%\end{wrapfigure}
\end{figure}

We first measure the amount of \emph{time to exhaust the search space} (TESS)
using the three versions of negations supported in \acheck, over a
bug-free version of the \sstlc benchmark for $n = 1, 2,\ldots$ up to the
point where we time-out. This gives some indication of how much of
the search space the three techniques explore, keeping in mind that
what is traversed is very different in shape; hence the
more reliable comparison is between \NE and \NEs. As the results
depicted in~\refFig{log_expl_stlc} suggests, \NEs shows a clear
improvement over \NE, while \NF holds its ground, however hindered by
the  explosive exhaustive  generation of terms. % More experiments of this
% sort are reported in~\cite{Pessina15}.  

%\begin{wrapfigure}{r}{0.6\textwidth}

However, our mission is finding counterexamples and so we compare the
\emph{time to find counterexamples} (TFCE) using \NF, \NE, \NEs on the said benchmarks.  We list in
Table~\ref{tab:stlc_tfce} the 9 mutations from the cited site.
%\url{http://docs.racket-lang.org/redex/benchmark.html}. 
Every row
describes the mutation inserted with an informal classification
inherited from ibidem --- (S)imple, (M)edium or (U)nusual, better read
as artificial. We also list the counterexamples found by \acheck under
\NF (NE(s) being analogous but less instantiated) and the depths at
which those are found or a time-out occurred.

%  Finally, as there is no support for logic programming,
% encoding non-algorithmic relations requires a fair amount of
% indirection.

% \begin{table} [tb]
%   \centering
%   \begin{tabular}{lll}
%     \toprule
%     bug\#&class&description\\ 
%     \midrule
%     1&S& app rule the range of the function is matched to the arg\\
%     2&M& \emph{(cons v) v} value has been omitted\\
%     3&S& swap of order of types in function pos of
%          app\\
%     5&S& tail reduction returns the head\\
%     6&M&hd reduction acts on partially applied cons\\
%     7&M&no evaluation RHS of app\\
%     8&U&lookup always returns int\\
%     9&S&vars may not match in lookup\\
%    \bottomrule
%   \end{tabular}
%  \caption{\sstlc benchmark bugs}
% \label{tab:stlc_bugs}
%  \end{table}

\begin{table}[tb]
  \centering
     \begin{tabular}{clllllll}
       \toprule
       bug&check&NF&NE&NEs&cex&Description/Class\\ 
       \midrule
       1&pres&0.3 (7)&1 (7)& 0.37 (7)&$(\lambda x.\ x \ err) \ n$&  range
                                                                of
                                                                function
                                                                   in
                                                                   app
                                                                   rule
                                                                                                                            \\
       & prog&0 (5)&3.31 (9)& 0.27 (5)&\emph{hd n}& \quad matched
                                                                to the
                                                                arg.~(S)\\
       2&prog&0.27 (8)&t.o. (11)&85.3 (12)&\emph{(cons n) nil}& value
                                                                \emph{(cons
                                                                v) v}
                                                                omitted
       (M)\\
       3&pres&0.04 (6)&0.04 (6)&0.3 (6)&$(\lambda x.\ n) \ m$& order
                                                               of
                                                               types
                                                               swapped\\
                                                               
       & prog&0 (5)&3.71 (9)& 0.27 (8)&\emph{hd n}&\quad in function pos of
          app (S)\\
       4 & prog& t.o.&t.o.&t.o.&?& the type of cons is incorrect (S) \\
       5&pres&t.o. (9)&t.o. (10)&41.5 (10)&\emph{tl ((cons n) err)}&
                                                                     tail
                                                                     red.\
                                                                     returns
                                                                     the
                                                                     head
       (S)\\
       6&prog&29.8 (11)&t.o. (11)& t.o. (12)&\emph{hd ((cons n)
                                              nil)}&hd red.\
                                                     on part.~applied cons (M)\\
       7&prog&1.04 (9)&18.5 (10)& 1.1 (9) & $hd\ ((\lambda x.\ err)
                                            \ n)$&no eval for
                                                   argument of app (M)\\
       8&pres&0.02 (5)&0.03 (5)&0.1 (5)&$(\lambda x.\  x) \
        nil$&lookup always returns int (U)\\
       9&pres&0 (5)&0.02 (5)&0.1 (5)&$(\lambda x.\  y) \ n$& vars do
                                                             not match
                                                             in lookup
       (S)\\ 
       \bottomrule
\\
     \end{tabular}
%   \end{center}
  % \begin{small}
    \caption{TFCE on the \sstlc benchmark, Redex-style encoding}
    \label{tab:stlc_tfce}
\end{table}

The results in Table~\ref{tab:stlc_tfce} show a remarkable improvement of \NEs over \NE, in
terms of counter-examples that were timed-out (bug 2 and 5), as well as major
speedups of more than an order of magnitude (bugs 3 (ii) and 7). Further,
\NEs never under-performs \NE, probably because it locates counterexample
at a lower depth. In rare occasions (bug 5 again) \NEs even
outperforms \NF and in several cases it is comparable (bug 1, 3, 7, 8
and 9). Of course there are occasions (2 and 6), where \NF is still
dominant, as \NEs counter-examples live at steeper depths (12 and 16,
respectively) that cannot yet be achieved within the time-out.

% \begin{table} [tb]
%   \begin{center}
%      \begin{tabular}{clllllll}
%        \toprule
%        bug\#&check&NF&NE&NEs&cex&notes\\ 
%        \midrule
%        1&pres&0.3 (7)&1 (7)& 0.37 (7)&$(\lambda x.\ x \ err) \ n$&\\
%        & prog&0 (5)&3.31 (9)& 0.27 (5)&\emph{hd n}&\\
%        2&prog&0.27 (8)&t.o. (11)&85.3 (12)&\emph{(cons n) nil}&NE: cex lives at
%                                                       depth 12\\
%        3&pres&0.04 (6)&0.04 (6)&0.3 (6)&$(\lambda x.\ n) \ m$&\\
%        & prog&0 (5)&3.71 (9)& 0.27 (8)&\emph{hd n}&\\
%        5&pres&t.o. (9)&t.o. (10)&41.5 (10)&\emph{tl ((cons n) err)}& NF: 4.1
%                                                             (10) with
%                                                             custom gen\\
%        6&prog&29.8 (11)&t.o. (11)& t.o. (12)&\emph{hd ((cons n) nil)}&NE(s): cex
%                                                            lives at
%                                                            depth 17(16)\\
%        7&prog&1.04 (9)&18.5 (10)& 1.1 (9) & $hd\ ((\lambda x.\ error) \ n)$&\\
%        8&pres&0.02 (5)&0.03 (5)&0.1 (5)&$(\lambda x.\  x) \
%         nil$&\\
%        9&pres&0 (5)&0.02 (5)&0.1 (5)&$(\lambda x.\  y) \ n$&\\ 
%        \bottomrule
%      \end{tabular}
%    \end{center}
%   % \begin{small}
%     \caption{\acheck results on the \sstlc benchmark}
%     \label{tab:stlc_tfce}
% \end{table}

We do not report TFCE of PLT-Redex, because, being based on randomized
testing, what we really should
measure is time spent \emph{on average} to find a
bug.  % which we could not reproduce
% reliably, but it seems very fast, in this benchmark under $1$
% second.
The two encodings are quite different: Redex has
very good support for evaluation contexts, while we use congruence
rules. Being untyped, the Redex encoding treats \emph{err} as a
string, which is then procedurally handled in the statement of
preservation and progress, whereas for us it is part of the
language. Since~\cite{Klein12}, Redex allows the user to write
certain judgments in a declarative style, provided they can be
given a functional mode, but more complex systems, such as typing
for a polymorphic version of a similar calculus, require very indirect
encoding, e.g.\ CPS-style. We simulate addition on integers with
numerals (omitted from the code snippets presented in
Section~\ref{sec:tut} for the sake of space), as we currently require
our code to be pure in the logical sense, %  i.e.\ no appeal to built-in
% arithmetic,
as opposed to Redex that maps integers to Racket's
ones. \emph{W.r.t.}~lines of code, the size of our encoding is roughly
$1/4$ of the Redex version, not counting Redex's built-in generators
and substitution function. The adopted checking philosophy is also
somewhat different: they choose to test preservation and progress
together, using a cascade of three built-in generators and collect all
the counterexamples found within a timeout.

The performance of the negation elimination variants in this benchmark
is not too impressive.
However, % we have reported those to
% demonstrate how NEs improves on NE/NF rather as a comparison to
% PLT-Redex. In fact,
if we adopt a different style of encoding (let's call it PCF, akin to
what we used in~\cite{CheneyM07}), where
constructors such as \texttt{hd} are \emph{not} treated as
constants, but are first class, e.g.:
\begin{small}
\begin{verbatim}
tc(G,hd(E),intTy)       :- tc(G,E,listTy).
step(hd(cons(H,Tl)), H) :- value(H),value(Tl).
\end{verbatim}
\end{small}
then all counter-examples are found very quickly, as reported in
Table~\ref{tab:minml_tfce}. In bug 4,  \NEs struggles to
get at depth 13:  on the other hand PLT-Redex
fails to find that very bug.  % Here we have also disposed of the
% \texttt{err} constant and added \texttt{plus}.
Bug 6 as well
as several counterexamples disappear as not
well-typed. This improved efficiency may be due to the reduced amount
of nesting of terms, which means lower depth of exhaustive
exploration. This is not a concern for random generation and
(compiled) functional execution as in PLT-Redex.% On the other hand
% this is payed in terms of increased number of clauses in the
% operational semantics, in particular encoding the congruence
% rules. This is probably why bug 7 cannot be found in the time limit.

\begin{table}[tb]
\centering
     \begin{tabular}{cllllll}
       \toprule
       bug\#&check&NF&NE&NEs&cex\\ 
       \midrule
       1&pres&0.05 (5)& 2.79 (5)& 0.04 (5)&$(\lambda x.\ hd \ x) \ N$\\
       2&prog&0 (4)& 7.76 (9)& 0.8 (7)&\emph{(cons N) nil}\\
                                                     
       3&pres&0 (4)& 0.05 (4)&0 (4)&$(\lambda x.\ nil) \ nil$\\
       4&prog&0.15 (7)&t.o. (10)&199.1 (12)&\textit{N + (cons N nil)}\\
       5&pres&0(4)& 0.04 (4)&0(4)&\emph{tl (cons N) nil}\\
  7&prog&5.82  (9)&151.2 (11)& 19.54. (10) & $(\lambda x.\ nil) \ ( N + M)$\\   
       %7???&prog&0  (5)&0.8 (6)& 0.1 (5) & $hd\ ((\lambda x.\ nil) \ N)$\\
       8&pres&0.01 (4)&0.04 (4)&0.1 (4)&$(\lambda x.\  x) \ nil$\\
       9&pres&0 (4)&0.04 (4)&0.1 (4)&$(\lambda x.\  y) \ N$\\ 
       \bottomrule\\
     \end{tabular}
  % \begin{small}
    \caption{TFCE on the
      \sstlc benchmark, PCF-style encoding. \NEs cex shown}\label{tab:minml_tfce}
\end{table}

\subsection{Nitpicking Security Type Systems}
\label{ssec:volpano}
To compare Nitpick with our approach, we use the security type system
due to Volpano, Irvine and Smith~\cite{Volpano1996}, whereby the basic
imperative language \emph{IMP} is endowed with a type system that
prevents information flow from private to public
variables\footnote{For an interesting case study regarding instead
  \emph{dynamic} information flow and carried out in Haskell,
  see~\cite{Hritcu}. A large part of the paper is dedicated to the
  fine tuning of custom generators and shrinkers.}. % Blanchette and
% Nipkow reports that, by inserting a mutation in the typing rule for
% command sequencing, they get a counterexample to the crucial
% \emph{non-interference} property.
For our test, we actually selected the more general version of the
type system formalized in~\cite{NipkowK14}, where the security levels
are generalized from \emph{high} and \emph{low} to natural
numbers. Given a fixed assignment \emph{sec} of such security levels
to variables, then lifted to arithmetic and Boolean expressions, the
typing judgment $l \vdash c$ reads as ``command $c$ does not contain
any information flow to variables $< l$ and only safe flows to
variables $\geq l$.'' Following~\cite{NipkowK14}, we call this system
\emph{syntax-directed}. % We show a selection of the rules, where the
% over-strike denotes the inserted mutation.

% \[
% \begin{array}{c}
% % \ianc{}{l \vdash SKIP}{}  
% % \quad
% \ibnc{\secl a\leq \secl x}{l \leq \secl x}{l \vdash x := a}{}
% \qquad
% \ibnc{l \vdash c_1}{\xcancel{l \vdash c_2}}{l \vdash c_1 ; c_2}{}
% \qquad
% \ibnc{max\ (\secl b \ l) \vdash c_1}{max\ (\secl b \ l) \vdash c_2}{l
%   \vdash IF \ b\ THEN \ c_1 \
% ELSE\ c_2}{}
% \end{array}
% \]
% \begin{metanote}
%   This can be shortened
% \end{metanote}
The main properties of interest relate states that agree on the value of
each variable (strictly) \emph{below} a certain security level,
denoted as $\siml{\sigma_1}{<}{\sigma_2}$ iff
$\forall x.\ \secl x < l \rightarrow \sigma_1(x) = \sigma_2(x)$. Assume
a standard big-step evaluation semantics for IMP, relating an initial
state $\sigma$ and a command $c$ to a final state $\tau$:
\begin{description}
\item[Confinement] If $\langle c,\sigma\rangle\downarrow \tau$ and $l\vdash
  c$ then $\siml{\sigma} {<} {\tau}$;
\item[Non-interference] If $\langle c,\sigma\rangle\downarrow
  \sigma'$,  $\langle c,\tau\rangle\downarrow \tau'$, 
  $\siml{\sigma} {\leq} {\tau}$ and  $0\vdash
  c$ then $\siml{\sigma'} {\leq} {\tau'}$;
\end{description}

We extend this exercise by considering also a \emph{declarative}
version (\emph{std})  $l \vdash_d c$ of the syntax directed system, where
anti-monotonicity is taken as a primitive rule instead of an admissible
one as in the previous system; finally we encode also a
syntax-directed \emph{termination-sensitive} (\emph{stT}) version
$l \vdash_{\Downarrow} c$, where non-terminating programs do not leak
information and its declarative cousin (\emph{stTd}) $l \vdash_{\Downarrow d} c$. We
then insert some mutations in all those systems, as detailed in
Table~\ref{tab:volp} and investigate whether the following equivalences
among those systems still hold:
\begin{description}
\item[st$\Bimp$std] $l \vdash c$ iff $l \vdash_d c$ and
  \textbf{stT$\Bimp$stTd} \ $l \vdash_{\Downarrow } c$ iff
  $l \vdash_{\Downarrow d} c$.
\end{description}

\begin{table}[tb]
   \centering
     \begin{tabular}{lllllll}
       \toprule
       bug&check&Nitpick&NF&NE&NEs&Description\\ 
       \midrule
       1&conf&(sp)&0.03 (5)&4.4 (8)&2.1 (7)& second premise of seq rule omitted\\
       & non-inter&t.o.&9.13 (8)&6.71 (8)&6.1 (8)& ditto \\
       2&non-inter&(sp)&3.3 (8)&2.1 (8)&1.9 (8)& var swapping in $\le$ premise of assn rule\\
       3&st$\rightarrow$std&0.95&t.o. & t.o &t.o. &inversion of $\le$ in
                                                    antimono rule
       \\
        &std$\rightarrow$st&0.75&0.8 (7)& 0.3 (7) &0.3 (7)&ditto\\
       4&st$\rightarrow$std & &&&& $\le$ assumption
                                                     omitted 
                                                      in IF: 
                          \textbf{true}\\
&std$\rightarrow$st&1.3&0.9 (7)&t.o. &t.o. & ditto \\
       5&st$\rightarrow$std&5.1(sp)&24.5  (11)&t.o.&t.o. &as 2 but on  decl.~version of the rule\\
        &std$\rightarrow$st&1.1& 0.2 (7)&t.o.& 24.6 (11) & ditto\\
       6&stT$\rightarrow$stTd&5.1(sp)&t.o.&t.o.&t.o.&as 2 but on term.~version of the rule\\
        &stTd$\rightarrow$stT&1.0&0.01 (5)&0.32 (7) &0.05 (6)&ditto
       \\
       7 &stT$\rightarrow$stTd&&&  &  &same as 4 but
                                                        on term-decl.~rule: \textbf{true} \\      
          &stTd$\rightarrow$stT&1.6&1.7 (8)&12.5 (9)&1.2(8) &  ditto\\
  \bottomrule\\
     \end{tabular}
  % \begin{small}
    \caption{\acheck vs.~Nitpick on the Volpano benchmark suite. (sp)
      indicates that Nitpick produced a spurious counterexample.}
    \label{tab:volp}
  % \end{small}
\end{table}

\begin{sloppypar}
  Again the experimental evidence is quite pleasing as far as \NE vs.\
  \NEs goes, where the latter is largely superior (5 (ii), 1 (i), 7
  (ii)). In one case \NEs improves on \NF (1 (ii)) and in general
  competes with it save for 4 (ii) and 5 (i) and (ii). To have an idea
  of the counterexamples found by \acheck, the command
  $\mathtt{(SKIP\ ; x := 1), \secl x = 0, l=1}$ and state $\sigma$ mapping $x$
  to $0$ falsifies confinement 1 (i); in fact, this would not hold were
  the typing rule to check the second premise. A not too dissimilar
  counterexample falsifies non-interference 1 (ii): $c$ is
  $\mathtt{(SKIP\ ; x := y), \secl x,y = 0,1, l=0}$ and $\sigma$ maps
  $y$ to $0$ and $x$ undefined (i.e.\ to a logic variable), while
  $\tau$ maps $y$ to $1$ and keeps $x$ undefined. We note in passing
  that here extensional quantification is indispensable, since
  ordinary generic quantification is unable to instantiate
security  levels so as to find the relevant bugs.
\end{sloppypar}

The comparison with Nitpick\footnote{Settings:
  \texttt{[sat\_solver=MiniSat\_JNI,max\_threads=1,timeout=200]}} is
more mixed. On one hand Nitpick fails to find 1 (ii) within the
timeout and in other four cases it reports \emph{spurious}
counterexamples, which on manual analysis turn out to be good.  On the
other it nails down, quite quickly, two other cases where \acheck
fails to converge at all (3 (i), 6 (i)). This despite the facts that
relations such as evaluations, $\vdash_d$ and $\vdash_{\Downarrow d}$,
are reported not well founded requiring therefore a problematic
unrolling.

The crux of the matter is that differently from Isabelle/HOL's
mostly functional setting (except for inductive definition of
evaluation and typing), our encoding is fully relational: states and
security assignments cannot be seen as partial functions but are
reified in association lists. Moreover, we pay a significant price in
not being able to rely on built-in types such as integers, but have to
deploy our clearly inefficient versions. This means that to falsify
simple computations such as $ n \le m$, we need to provide a
derivation for that failure.  Finally, this case study does not do
justice to the realm where \aprolog excels, namely it does not
exercise binders intensely: we are only using nominal techniques in
representing program variables as names and freshness to guarantee
well-formedness of states and of the table encoding the variable
security settings. Yet,  we could not select more binding intensive
examples due to the current
difficulties with running Nitpick under \emph{Nominal} Isabelle. 

% \begin{table}[h]
% \begin{tabular}{l|r|r|r|r}
% \textbf{Check} & \multicolumn{1}{l}{\textbf{Nitpick}} &
% \multicolumn{1}{l}{\textbf{NF}} & \multicolumn{1}{l}{\textbf{NE}} &
% \multicolumn{1}{l}{\textbf{NE$^-$}} \\
% \hline
% Confinement     & 0.6      & 0.04      & 0.04 & 0.03     \\
% Non-interference      &   timeout   & 9.46   & 0.32   & 0.29                               
% \end{tabular}
% \caption{The VIS case study: time out at 30 secs, max depth $5$ (for confinement) and
%   $8$ (for non-interference)}
% \label{tab:volp}
% \end{table}

%%% Local Variables:
%%% mode: latex
%%% TeX-master: "tap"
%%% End:

% \begin{table} [tb]
%   \begin{center}
%      \begin{tabular}{cllllll}
%        \toprule
%        bug\#&check&NF&NE&NEs&cex\\ 
%        \midrule
%        1&pres&0.02 (7)& 2.79 (5)& 0.28 (5)&$(\lambda x.\ hd \ x) \ N$\\
%        2&prog&0 (4)&0 (7)&0 (7)&\emph{(cons N) nil}\\
                                                     
%        3&pres&0 (4)& t.o. (9)&0 (4)&$(\lambda x.\ nil) \ nil$\\
%        4&prog&0.15 (7)&0.01 (9)&0.01 (9)&\textit{N + (cons N nil)}\\
%        5&pres&t.o. (9)& 0 (4)&0(10)&\emph{tl (cons N) nil)}\\
     
%        7&prog&0  (5)&0.8 (6)& 0.1 (5) & $hd\ ((\lambda x.\ nil) \ N)$\\
%        8&pres&0.02 (5)&0.03 (5)&0.1 (5)&$(\lambda x.\  x) \
%         nil$\\
%        9&pres&0 (5)&0.02 (5)&0.1 (5)&$(\lambda x.\  y) \ n$\\ 
%        \bottomrule
%      \end{tabular}
%    \end{center}
%   % \begin{small}
%     \caption{\acheck results on the \aprolog friendly version of the
%       same benchmark. Nes cex shown}
%     \label{tab:minml_tfce}
% \end{table}

%%% Local Variables:
%%% mode: latex
%%% TeX-master: "tap"
%%% End:

% -------------------
\section{Conclusions and Future Work}
\label{sec:concl}
 
We have presented a new implementation of the \NE algorithm underlying
our model checker \acheck and experimental evidence showing satisfying
improvements \wrt the previous incarnation, so as to make it
competitive with the \NF reference implementation. The comparison with
PLT-Redex and Nitpick, systems of considerable additional maturity, is
also, in our opinion, favourable: \acheck is able to find similar
counterexamples in comparable amounts of time; it is able to find some
counterexamples that Redex or Nitpick respectively do not; and in no
case does it report spurious counterexamples.  Having said that, our
comparison is at most just suggestive and certainly partial, as many
other proof assistants have incorporated some notion of PBT,
e.g.~\cite{Owre06randomtesting,QChick}. A notable absence here is a
comparison with what at first sight is a close relative, the Bedwyr
system~\cite{BaeldeGMNT07}, a logic programming engine that allows a
form of model checking directly on syntactic expressions possibly
containing binding.  Since Bedwyr uses depth-first search, checking
properties for infinite domains should be approximated by writing
logic programs encoding generators for a {finite} portion of that
model. Our initial experiments in encoding the \sstlc benchmark in
Bedwyr have failed to find any counterexample, but this could be
imputed simply to our lack of experience with the system.  Recent work
about ``augmented focusing systems''~\cite{HeathM15} could overcome
this problem.

All the mutations we have inserted so far have injected faults in the
specifications, not in the checks. This make sense for our intended
use; % , where the properties we validate are the main theorems that our
% calculi should satisfy.
however, it would be interesting to see how
our tool would fare \wrt mutation testing of \emph{theorems}.
%, for example using \emph{isabelle mutabelle}.

\emph{Exhaustive} term generation has served us well so far, but it is
natural to ask whether \emph{random} generation could have a role in
\acheck, either by simply randomizing term generation under \NF or
more generally the \lp interpreter itself, in the vein
of~\cite{FetscherCPHF15}. More practically, providing
generators and reflection mechanism for built-in datatypes and
associated operators is a priority.
%\begin{itemize}
% \item Something on random testing, Feat etc, generators for built-in
%   datatypes \dots
%\item 

Finally, we would like to implement improvements in nominal equational
unification algorithms, which would make subsumption complete, \via
\emph{equivariant} unification~\cite{cheney10jar}, and more ambitiously
introduce \emph{narrowing}, so that functions could be computed rather
then simulated relationally. In the long run, this could open the door
to use \acheck as a light-weight model checker for (a fragment) of
Nominal Isabelle.
%\end{itemize}

%%% Local Variables:
%%% mode: latex
%%% TeX-master: "tap"
%%% End:

% -------------------
\appendix

\section{Appendix}

\subsection{Some formal definitions}
\label{sec:app}

The effect of a permutation $\pi$ on a name:
\[\begin{array}{rcl}
\idd (\Aa) &=& \Aa\\
(\tran{\Aa}{\Ab} \comp \pi)(\Ac) &=& \left\{\begin{array}{ll}
\Ab & \pi(\Ac) = \Aa\\
\Aa & \pi(\Ac) = \Ab\\
\pi(\Ac) & \pi(\Ac) \notin \{\Aa,\Ab\}
\end{array}\right.
\end{array}\]
The swapping operation on \emph{ground} terms:
\[\begin{array}{rclcrclc}
  \pi \act \unit &= &\unit&\qquad&
  \pi \act f(t) &=& f(\pi \act t)\\
  \pi \act \pair{t,u} &=& \pair{\pi \act t, \pi \act u}& \qquad &
  \pi \act \Aa &= &\pi(\Aa)\\
  \pi \act \abs{\Aa}{t} &= &\abs{\pi \act \Aa}{\pi \act t}
\end{array}\]
Constraint satisfaction:
\begin{eqnarray*}
\theta \models \true \\
\theta \models t \eq u &\iff& \theta(t) \eq \theta(u)\\
\theta \models t \fresh u &\iff& \theta(t) \fresh \theta(u)\\
\theta \models C \andd C' &\iff& \theta \models C \textrm{ and
}\theta\models C'\\
\theta \models \exists X{:}\tau.~C &\iff& \textrm{for some $t:\tau$,
  $\theta[X:=t]\ednote{Undefined notation} \models C$}\\
\theta \models \new \Aa{:}\nu.~C &\iff& \textrm{for some $\Ab \fresh
  (\theta,C)$, $\theta \models C[\Ab/\Aa]$}
\end{eqnarray*}
A context $\Gamma$ is a sequence of bindings between variables (or
names) and types.
\[
\Gamma ::= \cdot \mid \Gamma,X{:}\tau \mid \Gamma\#\Aa{:}\nu
\]
where we write name-bindings as
$\Gamma\#\Aa{:}\nu$, to remind us that $\Aa$
must be fresh for other names and variables in $\Gamma$.

\smallskip
\noindent Term complementation:
 \begin{small}
  \begin{eqnarray*}
   \mnot\SB{\tau} &:&  \tau \to \tau\ \mathit{set}\\
    \mnot\SB\tau(t) & = & \emptyset\qquad\qquad\qquad\mbox{when } \tau\in\{
    \unitTy,\nu, \abs{\nu}\tau \} \mbox{ or $t$ is a variable}\\
    \mnot\SB{\tau_1 \times \tau_2}(t_1,t_2) & = & 
\{ (s_1,\_) \mid s_1\in\mnot\SB{\tau_1}(t_1)\}\cup\{ (\_,s_2) \mid
s_s\in\mnot\SB{\tau_2}(t_2)\}\\
\mnot\SB\delta(f(t)) & = & \{
g(\_)\mid g\in\Sigma, g \hastype \sigma\Imp \delta, f\ \neq g\} \cup \{f(s)\mid s\in\mnot\SB\tau(t) \}
  \end{eqnarray*}
\end{small} 

The correctness of the algorithm for
term complementation can be stated in the following
constraint-conscious way, as required by the proof of the main
soundness theorem:
\begin{lemma}[Term Exclusivity] 
\label{le:excluT}

Let $\constr$ be consistent, $ s \in \mnot\SB\tau(t)$,
$FV(u)\subseteq\Gamma$ and $FV(s,t)\subseteq\vec X$.  It is not the
case that both $\sat{\Gamma}{\constr}{\exists \vec X {:} \vec \tau.~u
  \eq t %\land \vec C
}$ and $\sat{\Gamma}{\constr}{\exists \vec X {:} \vec \tau.~u \eq
  s % \land \vec D
}$.
\end{lemma}
Inequality and non-freshness:

%\begin{figure}[h]
\begin{eqnarray*}
\neqt\SB{\tau} &:& \tau \times \tau \to o\\
\neqt\SB{\unitTy}(t,u) &=& \false\\
\neqt\SB{\tau_1 \times \tau_2}(t,u) &=& \neqt\SB{\tau_1}(\pi_1(t),\pi_1(u)) \orr \neqt\SB{\tau_2}(\pi_2(t),\pi_2(u))\\
\neqt\SB{\delta}(t,u) &=& \neqt_\delta(t,u)\\
\neqt\SB{\abs{\nu}{\tau}}(t,u) &=& \new \Aa{:}\nu.~\neqt\SB{\tau}(t \conc \Aa,u \conc \Aa)\\
\neqt\SB{\nu}(t,u) &=& t \fresh u\\
\neqt_\delta(t,u) &\ent& \bigvee\{\exists X,Y{:}\tau.~t \eq f(X) \andd
u \eq f(Y) \andd \neqt\SB{\tau}(X,Y) \\
&&\qquad\qquad\mid  f:\tau \to \delta \in \Sigma\}\\
&&\vee \bigvee\{\exists X{:}\tau,Y{:}\tau'.~t\eq f(X) \andd u \eq g(Y) \\
&&\qquad\qquad\mid  f:\tau \to \delta,g:\tau' \to \delta \in \Sigma,f
\not= g\}
\end{eqnarray*}
\begin{eqnarray*}
\nfr\SB{\nu,\tau} &:& \nu \times \tau \to o\\
\nfr\SB{\nu,\unitTy}(a,t) &=& \false\\
\nfr\SB{\nu,\tau_1 \times \tau_2}(a,t) &=&\nfr\SB{\nu,\tau_1}(a,\pi_1(t)) \orr \nfr\SB{\nu,\tau_2}(a,\pi_2(t))\\
\nfr\SB{\nu,\delta}(a,t) &=& \nfr_{\nu,\delta}(a,t)\\
\nfr\SB{\nu,\abs{\nu'}{\tau}}(a,t) &=& \new \Ab{:}\nu'.~\nfr\SB{\tau}(a,t \conc \Ab)\\
\nfr\SB{\nu,\nu}(a,b) &=& a \eq b\\
\nfr\SB{\nu,\nu'}(a,b) &=& \false \quad (\nu \neq \nu')\\
\nfr_{\nu,\delta}(a,t) &\ent& \bigvee\{\exists X{:}\tau.~t \eq f(X) \andd \nfr\SB{\nu,\tau}(a,X) \mid  f:\tau \to \delta \in \Sigma\}
\end{eqnarray*}
% \caption{Inequality and non-freshness}
% \label{fig:neq_nfr}
% \end{figure}

% \section{A sample negative program}

% We give here a listing of the \texttt{not\_tc} program after partial
% subsumption --- partial as two subsumed clauses for application slip in:
% \begin{small}
% \begin{verbatim}
% not_tc(_,lam(_),listTy).
% not_tc(_,lam(_),intTy).
% (new x:id. not_tc([(x,T)|G],M@x,U)) => not_tc(G,lam(M,T),funTy(T,U)).
% (forall T:ty. not_tc(G,M,funTy(T,U7)) ; not_tc(G,N,T)) => not_tc(G,app(M,N),U7).
% (forall T:ty. not_tc(G,M,funTy(T,listTy)) ; not_tc(G,N7,T)) => not_tc(G,app(M,N7),listTy).
% (forall T:ty. not_tc(G,M,funTy(T,intTy)) ; not_tc(G,N,T)) => not_tc(G,app(M,N),intTy).
% not_tc([],var(_),_).
% (neq_ty(T,T') ; fresh_id(X,X')), not_tc(G,var(X'),T') => not_tc([(X,T)|G],var(X'),T').
% neq_ty(tccf(C),T) => not_tc(_,c(C),T).
% \end{verbatim}
% \end{small}
\subsection{Other experiments}
\label{sec:quick}

Random testing has been present in Isabelle/HOL's
since~\cite{BerghoferN04} and has been recently enriched with a notion
of \emph{smart} test generators to improve its success rate w.r.t.\
conditional properties.  Exhaustive and symbolic testing follow the
SmallCheck approach~\cite{SmallCheck}. Notwithstanding all these improvements,
QuickCheck requires all code and specs to be \emph{executable} in the
underlying functional language, while % this be can handled in a
% logical framework such Isabelle/HOL,
many of the specifications that we are interested in are best seen as
\emph{partial} and \emph{not terminating}.

While not terribly exciting, these benchmarks, proposed and
measured in \cite{bulwahn2011smart} and taken from Isabelle
\emph{List.thy} theory are useful to set up a rough comparison with
Isabelle's QuickCheck. We show the checks in our logic programming
formulation, leaving to the reader the obvious meaning, noting only
that we use numerals as datatype.
\begin{small}
\begin{verbatim}
D1: distinct([X|XS]) => distinct(XS).
D2: distinct(XS),remove1(X,XS,YS) => distinct(YS).
D3: distinct(XS),distinct(YS),zip(XS,YS,ZS) => distinct(ZS).
S1: sorted(XS),remove_dupls(XS,YS) => sorted(YS).
S2: sorted(XS),insert(X,XS,YS) => sorted(YS).
S3: sorted(XS),length(XS,N),less_equal(I,J),less(J,N),
                            nth(I,XS,X),nth(J,XS,Y) => less_equal(X,Y).
\end{verbatim}
\end{small}

\begin{table}[t]
  \begin{footnotesize}
    \begin{tabular}{l l l l l l l l l l l l l l l l l l l}
\toprule
      &   & 9 & 10 & 11 & 12 & 13 & 14 & 15 & 16 & 17 & 18 & 19 &
      20 & 21 & 22 & 23 & 24 & 25 \\
\midrule

      \texttt{D1} & S  & 0 & 0 & 0 & 0.2 & 0.7 & 3.8 & 22 & 135 & 862 &  &  &  &  &  &  &  &  \\
      & \NF & 0 & 0  & 0 & 0 & 0 & 0 & 0 & 0 & 0 & 0.07 & 0.12 & 0.2 & 0.32 & 0.52 & 0.83 & 1.36 & 2.22 \\
      & \NE & 0  & 0 & 0 & 0 & 0 & 0 & 0 & 0 & 0 & 0.06 & 0.11 & 0.18 & 0.3 & 0.49 & 1.8 & 1.3 & 2.1 \\
      & \NEs & 0  & 0 & 0 & 0 & 0 & 0 & 0 & 0 & 0 & 0.06 & 0.11 & 0.18 & 0.3 & 0.4 & 0.6 & 1.0 & 1.7 \\
\midrule

      \texttt{D2} & S &  0 & 0 & 0.1 & 0.4 & 2.5 & 16 & 98 & 671 &  &  &  &  &  &  &  &  &  \\
      & \NF & 0 & 0 &  0 & 0 & 0 & 0 & 0 & 0 & 0 & 0 & 0.07 & 0.19 & 0.32 & 0.51 & 0.83 & 1.36 & 2.23 \\
      & \NE & 0 & 0  & 0 & 0 & 0 & 0 & 0 & 0 & 0 & 0.6 & 0.11 & 0.18 &
                                                                      0.3 & 0.49 & 0.8 & 1.32 & 2.17 \\
      & \NEs & 0 & 0  & 0 & 0 & 0 & 0 & 0 & 0 & 0 & 0.6 & 0.11 & 0.18 &
                                                                      0.2 & 0.39 & 0.6 & 1.1 & 1.7 \\
\midrule

      \texttt{D3} & S &  4.3 & 157 &  &  &  &  &  &  &  &  &  &  &  &  &  &  &  \\
      & \NF & 0 & 0 &  0 & 0.08 & 0.14 & 0.35 & 0.76 & 1 & 3 & 6 & 12 & 24 & 45 & 82 & 155 & 286 & 580 \\
      & \NE & 0 & 0  & 0 & 0.08 & 0.13 & 0.32 & 0.68 & 1.3 & 3 & 6 & 11
                                                                & 22 &
                                                                       42 & 79 & 150 & 280 & 586 \\
      & \NEs & 0 & 0  & 0 & 0.08 & 0.13 & 0.22 & 0.5 & 0.9 & 2.1 & 4.5 & 8
                                                                & 17 &
                                                                       3 & 63 & 121 & 225 & 448 \\
\midrule

      \texttt{S1} & S &  0 & 0 & 0 & 0 & 0 & 0 & 0 & 0 & 0.10 & 0.2 & 0.3 & 0.8 & 1.7 & 3.6 & 7.8 & 17 & 36 \\
      & \NF & 0 & 0 &  0 & 0 & 0 & 0 & 0 & 0 & 0 & 0 & 0.6 & 0.08 & 0.11 & 0.15 & 0.21 & 0.27 & 0.35 \\
      & \NE & 0 & 0  & 0 & 0 & 0 & 0 & 0 & 0 & 0 & 0 & 0.06 & 0.08 &
                                                                    0.11 & 0.15 & 0.2 & 0.27 & 0.36 \\
      & \NEs& 0 & 0  & 0 & 0 & 0 & 0 & 0 & 0 & 0 & 0 & 0    & 0.04 &
                                                                    0.06 & 0.08 & 0.11&0.16 & 0.2 \\
\midrule

      \texttt{S2} & S &  0 & 0 & 0 & 0 & 0 & 0.1 & 0.1 & 0.2 & 0.5 & 1.1 & 2.5 & 5.5 & 12 & 28 & 61 & 135 & 292 \\
      & \NF & 0 & 0 &  0 & 0 & 0 & 0 & 0 & 0 & 0 & 0 & 0 & 0.05 & 0.07 & 0.1 & 0.13 & 0.18 & 0.23 \\
      & \NE & 0 & 0  & 0 & 0 & 0 & 0 & 0 & 0 & 0 & 0.06 & 0.08 & 0.11 &
                                                                       0.15 & 0.19 & 0.25 & 0.33 & 0.44 \\
      & \NEs& 0 & 0  & 0 & 0 & 0 & 0 & 0 & 0 & 0 & 0.02 & 0.04 & 0.04 &
                                                                       0.06 & 0.08 & 0.11 & 0.16 & 0.2  \\
\midrule

      \texttt{S3} & S &  0 & 0 & 0 & 0 & 0.1 & 0.1 & 0.2 & 0.4 & 0.9 & 2.2 & 5.1 & 12 & 26 & 59 & 136 & 311 & 708 \\
      & \NF & 0 & 0 &  0.05 & 0.08 & 0.13 & 0.2 & 0.32 & 0.48 & 0.73 & 1 & 1.5 & 2.2 & 3.2 & 4.5 & 6.4 & 8.9 & 12 \\
      & \NE & 0  &  0 & 0 & 0.05 & 0.08 & 0.12 & 0.18 & 0.27 & 0.4 &
                                                                    0.57 & 0.83 & 1.1 & 1.6 & 2.2 & 3.2 & 4.3 & 5.7\\
      & \NEs& 0  &  0 & 0 & 0    & 0    & 0    & 0.04 & 0.09 & 0.1 &
                                                                    0.28 & 0.4  & 0.5 & 0.8 & 1.1 & 1.5 & 2.1 & 2.9\\
   \bottomrule
    \end{tabular}
  \end{footnotesize}
\label{tab:lukas}
    \caption{TESS for list benchmark.}
 \end{table}

 Table~\ref{tab:lukas} shows the TESS run time up to a given size
 ($25$), that in our case we interpret as depth-bound. We extrapolated
 from Table 2 in \cite{bulwahn2011smart} the \emph{S} (for \emph{smart
   generator}) rows.  We omit the results for \emph{exhaustive} and
 \emph{narrowing-based} testing; the point of their inclusion was to
 show how smart generation outperforms the latter two over checks with
 hard-to-satisfy premises. Again, these measurements are only
 suggestive, since QuickCheck's result are taken with another hardware
 (empty cells denote timeout after $1$h as
 in~\cite{bulwahn2011smart}'s setup).  Still,  we are 
 largely superior, possibly due to smart generation trying to
 replicate in a functional setting what logic programming naturally
 offers.  Note however that tests in Isabelle/QuickCheck are
 efficiently run by code generation at the ML level, while our bounded
 solver is just a non-optimized logic programming interpreter -- to
 name one, it does not have yet first-argument indexing.

 As usual in TESS, negation elimination tends to outperform \NF,
 especially when, as here, it does not require extensional
 quantification. \NEs only marginally improves on \NE, because the
 negated predicates (\texttt{distinct,sorted} etc.) are already
 quite simple. % and does not pay the price involved with

%%% Local Variables:
%%% mode: latex
%%% TeX-master: "long-tap.tex"
%%% End:

% % -------------------

\bibliographystyle{abbrv}
\bibliography{../mmm-jv/mmm,tap}

\end{document}